\documentclass[aps,prd,twocolumn,superscriptaddress,nofootinbib,longbibliography,floatfix]{revtex4-2}

\usepackage[utf8]{inputenc}
\usepackage[T1]{fontenc}
\usepackage{amsmath,amssymb,bm}
\usepackage{mathrsfs}
\usepackage{graphicx}
\usepackage{booktabs}
\usepackage{array}
\usepackage[colorlinks=true,linkcolor=blue,citecolor=blue,urlcolor=blue]{hyperref}

\newcommand{\dd}{\mathrm{d}}
\newcommand{\Lag}{\mathcal{L}}

\newcommand{\calS}{\mathcal{S}}

\newcommand{\calD}{\mathcal{D}}
\newcommand{\Mp}{M_{\rm p}}
\newcommand{\Lam}{\Lambda}

\newcommand{\DES}{\mathrm{DES}}
\newcommand{\DESI}{\mathrm{DESI}}
\newcommand{\SN}{\mathrm{SN}}
\newcommand{\BAO}{\mathrm{BAO}}
\newcommand{\varphione}{\varphi_1}
\newcommand{\varphitwo}{\varphi_2}
\newcommand{\chiT}{\chi^2_{\rm tot}}

\begin{document}

\preprint{AI-MG-v5.2.3-aic}

\title{\texttt{CosmoAgent}: AI Agent for Theoretical Cosmology}

\author{Stephon Alexander}
\email{stephon\_alexander@brown.edu}
\author{Loukas Gouskos}
\email{loukas\_gouskos@brown.edu}
\author{Savvas M. Koushiappas}
\email{savvas\_koushiappas@brown.edu}
\affiliation{Department of Physics, Brown University, Providence, Rhode Island 02912, USA}
\affiliation{Brown Center for Theoretical Physics and Innovation (BCTPI), Providence, Rhode Island 02912, USA}

\author{Cooper Niu}
\email{cooper@primordiant.com}
\affiliation{Department of Physics, Brown University, Providence, Rhode Island 02912, USA}
\affiliation{Brown Center for Theoretical Physics and Innovation (BCTPI), Providence, Rhode Island 02912, USA}
\affiliation{Primordiant, San Francisco, CA 94124, USA}

\author{Wenrong Sun}
\email{wenrong\_sun@brown.edu}
\affiliation{Department of Physics, Brown University, Providence, Rhode Island 02912, USA}
\affiliation{Brown Center for Theoretical Physics and Innovation (BCTPI), Providence, Rhode Island 02912, USA}
\date{\today}
\date{\today}

\begin{abstract}
We present \texttt{CosmoAgent}, an agentic AI framework using large language models (LLMs) to search for dark-energy and modified-gravity models. We define a covariant grammar that allows LLM agents to explore the space of modified-gravity Lagrangians. An in-house computational pipeline derives the equations of motion, solves the background dynamics, and evaluates the likelihood using late-time expansion data, including the DES 5-year type Ia supernova sample and the DESI Data Release~2 BAO measurements. Theories are ranked by the Akaike Information Criterion (AIC) to balance simplicity and likelihood. We benchmarked $8$ frontier LLMs from ChatGPT, Claude, Gemini, Kimi, DeepSeek, and Qwen, and all LLMs are able to find theories with a lower nominal AIC than flat $\Lambda$CDM, with a best improvement of $\Delta{\rm AIC}=-3.328$. Seven recover quadratic quintessence. Exploratory searches also return induced-gravity and scalar--Gauss--Bonnet structures. The modest AIC improvements do not establish an observational preference for these models over flat $\Lambda$CDM. The code is publicly available at \href{https://github.com/wsunamoutlook-prog/AI-Modified-Gravity}{wsunamoutlook-prog/AI-Modified-Gravity}.
\end{abstract}

\maketitle
\raggedbottom

\section{Introduction}
\label{sec:introduction}
The energy source of late-time cosmic acceleration remains an open problem
\cite{Riess1998,Perlmutter1999}. $\Lambda$CDM provides
a remarkable and minimal description of the background expansion
\cite{Planck2018VI}, but current distance measurements do not determine
whether the acceleration arises from a cosmological constant,
dynamical fields, or modified gravity
\cite{Clifton2012,Kobayashi2019}. The DES supernova and DESI Data
Release 2 BAO measurements provide precise late-time data with which
to investigate these possibilities \cite{DES2024SN,DESI2025BAO}.

A wide range of theories has been proposed to explain late-time
cosmic acceleration~\cite{Clifton2012}. Among these possibilities, quintessence provides a well-studied framework for dynamical dark energy \cite{RatraPeebles1988}
and has received attention following the recent DESI results \cite{Akrami2025Quintessence}. Other approaches include ghost condensation
\cite{ArkaniHamed2004Ghost}, $f(R)$ gravity
\cite{HuSawicki2007}, and scalar--Gauss--Bonnet models
\cite{NojiriOdintsovSasaki2005}, which explore different scalar dynamics
and modifications of the gravitational action. A systematic search through this theory space is therefore important
for identifying viable explanations of the observed late-time expansion. The revolution in generative and agentic AI offers new tools for such systematic search.
Recent examples in particle physics include FERMIACC for generating and testing theoretical hypotheses~\cite{Agrawal2026Fermiacc},
ALBERT for grammar-constrained model discovery~\cite{Alexander2026Albert},
and \texttt{llm4smeft} for proposing operator combinations in the
Standard Model effective field theory~\cite{Hammad2026SMEFT}. 
LLMs coupled to computer algebra have also been used to identify
propagating degrees of freedom in cosmological perturbations, with
worked examples supplied through in-context learning~\cite{HellThiele2026AlgorithmicPhysics}.

Automated model discovery in cosmology has a substantial history that predates
language models. Genetic algorithms have been used to reconstruct the dark-energy
equation of state and the expansion history without assuming a parametric form
\cite{Nesseris:2012GA, Castelao:2026CosmoGen}, while symbolic regression searches
the space of closed-form expressions directly \cite{Udrescu:2020AIF}; its
exhaustive variant enumerates candidate expressions up to a fixed complexity and
ranks them by minimum description length \cite{Bartlett:2024ESR, Desmond:2026MDL},
a criterion that, unlike AIC or BIC, can also price the cost of selecting a model
from within a large candidate set. These approaches search a space of functions
fitted directly to data. The search presented here differs in its object:
candidates are covariant actions, from which the background equations are derived
symbolically and filtered for derivative order before any likelihood is evaluated.
The output is therefore {\it a Lagrangian with a validated background solution
rather than a fitted curve}. Roughly two-thirds of grammar-legal proposals are
rejected by the background and numerical checks (Sec.~\ref{subsec:search-controls}).
In this work, we present \texttt{CosmoAgent}, a proof-of-concept framework
for exploring modified-gravity theories as possible explanations
of late-time cosmic expansion.

A common concern when using LLMs for theoretical research is hallucination: they may propose inconsistent initial conditions, make errors in symbolic calculations, or draw unsupported scientific conclusions. 
Studies of LLM-assisted research have documented citation errors in
literature reviews~\cite{Hell2026LiteratureReview} and biases in automated
proposal assessment~\cite{Liu2026ProjectPlanning}.
To constrain hallucinations, we design a harness around the LLM
agents using LangGraph, a framework for coordinating agent workflows~\cite{LangGraph2024}, to ensure that the results are grounded in deterministic calculations and observational data. Moreover, distinct Lagrangian expressions need not represent
physically distinct theories: actions related by boundary terms or invertible field redefinitions can describe equivalent dynamics~\cite{Arzt1995}. We examine such equivalences among the recovered representatives. During the search, candidates are ranked using the Akaike Information Criterion (AIC), which balances fit quality against a penalty for the nominal number of free parameters~\cite{Akaike1974,Liddle2007}.

We benchmark eight LLMs (GPT-5.6 Sol, Claude Sonnet 5, Claude Opus 5, Claude Fable 5, Gemini 3.6 Flash, Qwen 3.7 Max, DeepSeek V4 Pro, and Kimi K3) with each agent evaluating 1000 initial theory proposals under a common search protocol. All eight LLMs identify candidate theories with lower nominal AIC than flat $\Lambda$CDM, and seven recover models belonging to the quadratic-quintessence basin. The searches also identify other structures of interest, including
axion-like scalar models, induced-gravity models, and scalar--Gauss--Bonnet couplings.
This proof-of-concept study focuses on cosmological background
evolution; a comprehensive perturbative stability analysis and Solar System tests
of the recovered models are left for future work.

\section{Theory Space}
\label{sec:theory-space}

\subsection{Dimension-balanced effective Lagrangians}
\label{subsec:dimension-balanced}

We organize the search space by operator mass dimension, following
the effective-field-theory approach to gravity~\cite{Burgess2004}.
A generated theory is stored and validated as a Lagrangian density,
\begin{equation}
\calS=\int \dd^4x\,\mathscr{L},
\qquad
\mathscr{L}=\sqrt{-g}\,(\Lag+\mathcal L_m) ,
\label{eq:action}
\end{equation}
with the scalar Lagrangian
\begin{equation}
\mathcal{L}
=
M_p^2
\sum_{I\in G\cup N}
c_I\Lambda^{2-d_I}\mathcal{O}_I
+
\sum_{I\in S}
c_I\Lambda^{4-d_I}\mathcal{O}_I
+
\mathcal{L}_{\rm scalar}
\label{eq:lagrangian}
\end{equation}

Here $G$, $N$, and $S$ are mutually disjoint index sets for
pure-curvature, nonminimal scalar--curvature, and generated pure-scalar
operators, respectively. For every $I\in G\cup N\cup S$,
$\mathcal{O}_I$ denotes the corresponding operator, $d_I$ its mass
dimension, and $c_I$ its dimensionless coefficient. The operator sector
determines the power of $\Lambda$ and whether the term carries the
$M_p^2$ prefactor, while $\mathcal{L}_{\rm scalar}$ contains the fixed
kinetic and potential terms of the active scalars. The ordinary-matter contribution is denoted by $\mathcal L_m$. We use the reduced Planck mass,
$M_p\equiv(8\pi G)^{-1/2}=2.435\times10^{18}\,\mathrm{GeV}$.

The operator building blocks are
\begin{equation}
R,\qquad
R_{\mu\nu},\qquad
R^{\rho}{}_{\sigma\mu\nu},\qquad
\nabla_{\mu},\qquad
\phi_{1},\qquad
\phi_{2}.
\end{equation}
Their covariant contractions generate pure-curvature, scalar, and
mixed scalar--curvature operators through mass dimension six. After
imposing the algebraic index symmetries and removing expressions that
are pairwise proportional under direct evaluation on a spatially flat
FLRW background, the implemented grammar retains a finite dictionary
of 141 candidate operators. This deliberately redundant dictionary defines the grammar-restricted
proposal space explored here. Its 141 entries are neither an independent
basis nor a completeness claim for all covariant operators through mass
dimension six. The full candidate catalog
is listed in Appendix~\ref{sec:Curvature invariants},
Tables~\ref{tab:operators-d2-d4}, \ref{tab:operators-d5}, and
\ref{tab:operators-d6}.

The scalar configurations are 
\begin{equation}
\hbox{none},\qquad \phi_1\hbox{ only},\qquad \phi_2\hbox{ only},\qquad \phi_1\text{ and }\phi_2 .
\end{equation}
When a scalar is active, the fixed scalar sector contains a canonical kinetic term and either a quadratic or an axion-like potential:
\begin{align}
\Lag_{\rm scalar}^{(i)}
&=-\frac12 g^{\mu\nu}\nabla_\mu\phi_i\nabla_\nu\phi_i - V_i(\phi_i),\\
V_i(\phi_i)
&=\frac12 m_i^2\phi_i^2
\quad{\rm or}\quad
m_i^2f_i^2\left[1-\cos\left(\frac{\phi_i}{f_i}\right)\right].
\label{eq:fixed-sector}
\end{align}
The quadratic branch is the standard canonical quintessence form,
whereas the periodic branch is the pseudo-Nambu--Goldstone-boson
realization of dynamical dark energy
\cite{RatraPeebles1988,Frieman1995,CaldwellLinder2005}.

The two scalar fields play different roles in the current theory grammar.  The first scalar is naturally measured in Planck units,
\begin{equation}
\phi_1=\Mp\varphione,
\end{equation}
while the second is measured in cutoff units,
\begin{equation}
\phi_2=\Lam\varphitwo .
\end{equation}
The hierarchy allows the search to include both Planck-normalized scalar-tensor sectors and cutoff-normalized EFT scalar structures.  The distinction matters because a term such as $\phi_2 R$ carries a different scale from $\phi_1 R$, even when their dimensionless field amplitudes are comparable.  It is therefore useful to parameterize the EFT cutoff in units of the present Hubble scale,
\begin{equation}
\lambda_H\equiv\frac{\Lam}{H_0}.
\label{eq:lambdaH}
\end{equation}
With $h\equiv H/H_0$, the dimensionless ratio between the background expansion scale and the cutoff is simply $H/\Lam=h/\lambda_H$.  This quantity is monitored by the validation-stage scale-separation cut.

For the ordinary-matter sector, the present analysis assumes a homogeneous pressureless-dust background, so its microscopic Lagrangian is not specified explicitly. Only its contribution to the background equations of motion is required. We assume minimal coupling and covariant conservation,
\begin{equation}
\nabla_\mu T_{(m)}^{\mu\nu}=0,
\qquad
T_{\mu\nu}^{(m)}=\rho_m u_\mu u_\nu,
\qquad
p_m=0,
\end{equation}
which on the FLRW background gives
\begin{equation}
\dot\rho_m+3H\rho_m=0,
\qquad
\rho_m(a)=\rho_{m0}a^{-3}.
\end{equation}
We parameterize the present matter density by
\begin{equation}
\Omega_{m0}\equiv
\frac{\rho_{m0}}{3M_p^2H_0^2},
\qquad
\rho_m(a)=3M_p^2H_0^2\Omega_{m0}a^{-3},
\end{equation}
where $\Omega_{m0}$ is sampled as a free parameter. Since
minimally coupled dust introduces no higher-time derivatives,
its source is added after the background derivative-order reduction.

\section{Mini-Superspace and Background Physical Filtering}
\label{sec:ghost}

\subsection{Background equations}
\label{subsec:background-eom}

The first physical filter is applied after reducing the covariant theory on the spatially flat FLRW metric
\begin{equation}
\dd s^2=-N(t)^2\dd t^2+a(t)^2\delta_{ij}\dd x^i\dd x^j .
\label{eq:flrw}
\end{equation}
The minisuperspace variation gives
\begin{equation}
\delta \calS=\int \dd t\left(E_N\delta N+E_a\delta a+\sum_iE_{\phi_i}\delta\phi_i\right),
\label{eq:variation}
\end{equation}

where the inactive scalar equations are omitted.  After variation, the equations are evaluated in cosmic-time gauge, $N=1$:

\begin{align}
&E_N
=E_N^{(g+\phi)}-3M_p^2H_0^2\Omega_{m0}=0,
\qquad\\
&E_a=E_a^{(g+\phi)}=0,
\qquad
E_{\phi_i}=E_{\phi_i}^{(g+\phi)}=0.
\label{eq:eom-set}
\end{align}
Here $E_N$ gives the matter-sourced modified Friedmann
constraint, while $E_a$ and $E_{\phi_i}$ determine the
dynamical evolution. Since $E_N$ is the unnormalized
minisuperspace lapse equation, dividing it by $a^3$ recovers
the standard dust scaling $\rho_m\propto a^{-3}$.

\subsection{Equations of motion and background derivative-order filtering}
\label{subsec:ostro}

Higher-curvature and derivative scalar--curvature operators can introduce independent third- or higher-order time derivatives into the FLRW equations, which can lead to an Ostrogradsky instability (ghost) in nondegenerate theories \cite{Ostrogradsky1850, Woodard2015}. The pipeline therefore applies a FLRW background derivative-order filter before likelihood evaluation.

The proposal grammar is not restricted to the covariant Horndeski class. Within its finite mass-dimension and derivative truncation, it also includes operators such as $R^2$, $R^3$, $\nabla_\mu R\nabla^\mu R$, and mixed scalar--curvature contractions. After reduction to FLRW, the EOM stage derives the corresponding minisuperspace equations and identifies their third- and higher-order time-derivative terms, which can be written schematically as
\begin{align}
E_A
&=E_A^{\leq 2}+
\sum_B \sum_{r\geq 3}
{\cal A}^{(r)}_{AB}
\nonumber\\[-2pt]
&\quad\times
\!\left(N,a,\dot N,\dot a,\phi_i,\dot\phi_i,\ldots;c_I\right)
\,\frac{d^{r}y_B}{dt^r} ,
\label{eq:forbidden-derivatives}
\end{align}
where $A$ labels the lapse, scale-factor, and active-scalar equations, $y_B\in\{N,a,\phi_i\}$, and ${\cal A}_{AB}^{(r)}$ depends algebraically on the operator coefficients and lower-order variables. The background derivative-order filter imposes
\begin{equation}
{\cal A}_{AB}^{(r)}=0
\qquad
\hbox{for all independent }B\hbox{ and }r\geq 3,
\label{eq:cancellation-relations}
\end{equation}
and substitutes the resulting coefficient relations before gauge fixing and numerical construction. 

Equation~\eqref{eq:cancellation-relations} eliminates the identified
third- and higher-order time derivatives, leaving retained FLRW equations
of at most second order. This is a background selection criterion:
it neither establishes positivity of the kinetic and gradient terms of
propagating modes nor includes every healthy degenerate theory
\cite{Horndeski1974,Gleyzes2015BeyondHorndeski,Langlois2016DHOST}.
Covariant and perturbative stability therefore require separate checks.
Lagrangian methods for identifying constraints and counting propagating
degrees of freedom provide a starting point for such
analyses~\cite{Hell2026DegreesOfFreedom}.

\subsection{The modified Friedmann constraint}
\label{sec:hamiltonian}
\label{subsec:modified-friedmann}

In GR, the FLRW lapse $N$ is a Lagrange multiplier whose equation gives the Friedmann constraint, the Hamiltonian constraint of the homogeneous system. In a raw modified-gravity minisuperspace Lagrangian, however, derivatives of $N$ may appear before integrations by parts or before imposing the FLRW derivative-cancellation relations. We therefore identify the constraint directly from residual time-reparametrization invariance.
The use of the lapse as a constraint variable follows the standard ADM
decomposition of general relativity \cite{ADM1962}.

Under $t\rightarrow t-\xi(t)$, the minisuperspace variables transform as
\begin{equation}
\delta_\xi a=\xi\dot a,\qquad
\delta_\xi\phi_i=\xi\dot\phi_i,\qquad
\delta_\xi N=(N\xi)^{\cdot}.
\label{eq:time-diffeo}
\end{equation}
Invariance of the action gives the off-shell Noether identity
\begin{equation}
N\dot E_N-\dot a\,E_a-\sum_i\dot\phi_i\,E_{\phi_i}=0,
\label{eq:noether}
\end{equation}
which is the minisuperspace form of the Bianchi identity. After fixing $N=1$, it becomes
\begin{equation}
\dot E_N=\dot a\,E_a+\sum_i\dot\phi_i\,E_{\phi_i}.
\label{eq:noether-gauge}
\end{equation}

After the cancellation relations in Eq.~\eqref{eq:cancellation-relations} are imposed, the dynamical equations $E_a$ and $E_{\phi_i}$ are at most second order. The right-hand side of Eq.~\eqref{eq:noether-gauge} therefore contains no independent third derivatives. Because the identity holds off shell for arbitrary field profiles, the gauge-fixed lapse equation cannot contain independent accelerations; otherwise $\dot E_N$ would generate unmatched third derivatives. Hence it has the form
\begin{equation}
E_N\big|_{N=1}=C_N(a,\dot a,\phi_i,\dot\phi_i),
\label{eq:first-order-constraint}
\end{equation}
and is a first-order modified Friedmann constraint rather than an additional dynamical equation.

Equation~\eqref{eq:noether-gauge} also guarantees constraint propagation. On solutions of $E_a=E_{\phi_i}=0$, one has $\dot E_N=0$; therefore imposing $E_N(t_0)=0$ ensures $E_N(t)=0$ throughout the background evolution. This justifies using the lapse equation to determine one initial velocity and monitoring its residual during numerical integration. Thus, within the FLRW derivative-cancelled sector, the modified Friedmann constraint consistently removes the redundant initial datum and is preserved by the dynamical equations.

\section{Cosmological Observables}
\label{sec:observables}

Each accepted solution gives a Hubble history $H(z)=H_0h(z)$.  The supernova distance modulus is computed from the luminosity distance~\cite{Hogg1999}
\begin{equation}
D_L(z)=(1+z_{\rm hel})\frac{c}{H_0}
\int_0^{z_{\rm HD}}\frac{\dd z'}{h(z')},
\label{eq:luminosity-distance}
\end{equation}
with an analytically profiled absolute-magnitude offset. We use the
1820-object DES-Dovekie recalibration of the DES-SN5YR sample
\cite{DES2024SN,Popovic2026Dovekie}, denoted DES-SN throughout.
This likelihood therefore mainly constrains the relative shape of the
distance-redshift relation.

The BAO observables are computed from the same background solution:
\begin{align}
\calD_M(z)&=\int_0^z\frac{\dd z'}{h(z')},\\
\calD_H(z)&=\frac{1}{h(z)},\\
\calD_V(z)&=\left[\frac{z\calD_M(z)^2}{h(z)}\right]^{1/3}.
\label{eq:bao}
\end{align}
These are the standard transverse, radial, and volume-averaged BAO
distance combinations \cite{Eisenstein2005}. In this background-only analysis, the BAO scale is profiled as a nuisance parameter.  Thus the comparison with DESI DR2 BAO data~\cite{DESI2025BAO} should be interpreted as a relative-distance test of the late-time expansion history, not as a complete early-universe calculation of the sound horizon in each modified-gravity theory.

The likelihood statistic retained for every numerical sample is
\begin{equation}
\chiT=\chi^2_{\DES\text{-}\SN}+\chi^2_{\DESI\text{-DR2}\,\BAO}.
\label{eq:chi2}
\end{equation}
This statistic supplies the fit information used by the parameter sampler and the cross-structure objective described in Sec.~\ref{subsec:aic-objective}.

\section{LLM-Guided Search}
\label{sec:search}
\begin{figure*}[t]
\centering
\includegraphics[width=0.88\textwidth]{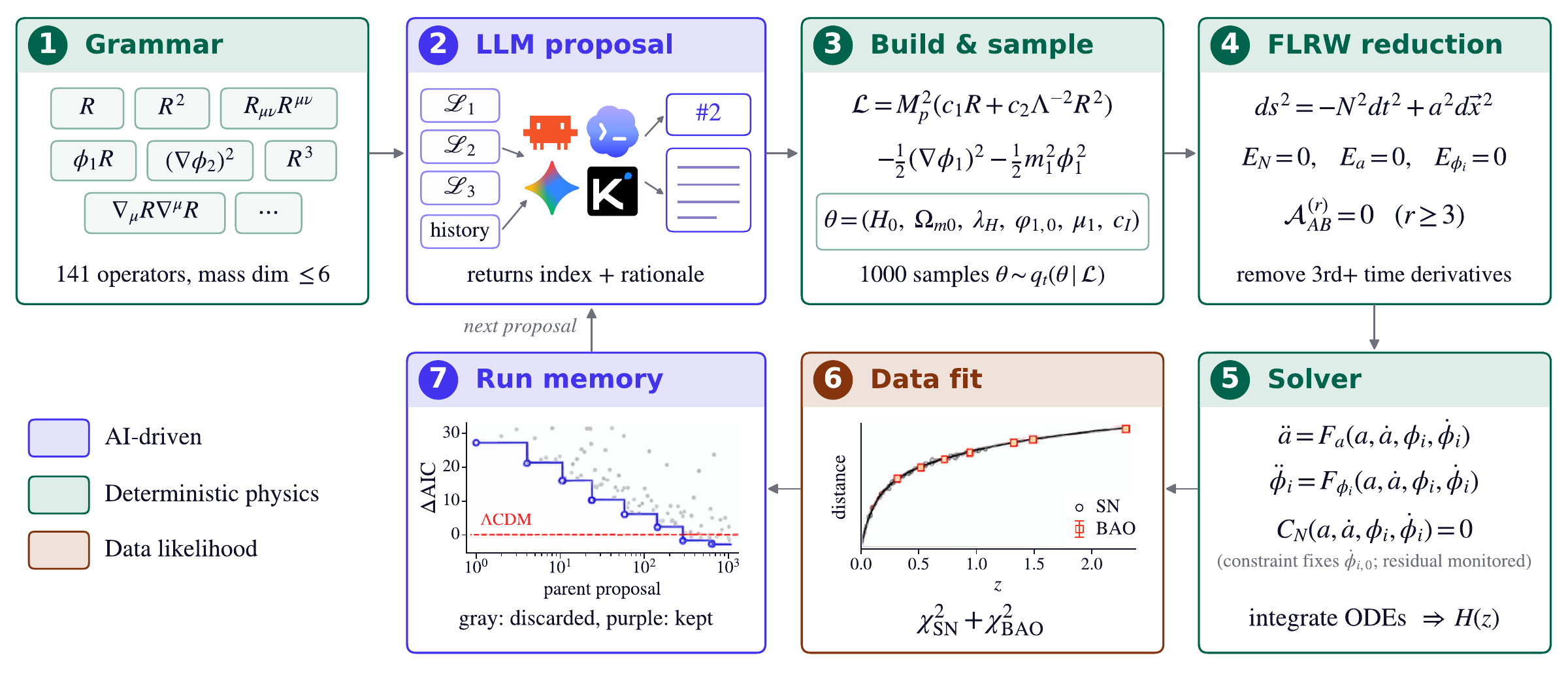}
\caption{Architecture of the autonomous search.  Blue modules form the AI and LangGraph orchestration layer; green modules are deterministic physics stages; and the red module evaluates the DES-SN plus DESI-DR2 BAO likelihood.  AIC compares distinct structures, while $\chi^2_{\rm tot}$ ranks parameter samples within one structure.  Feedback changes where the agent searches but cannot override validation. The LLM guides candidate selection, while symbolic derivation, numerical fitting, and the implemented background checks are handled by the deterministic pipeline.}
\label{fig:workflow}
\end{figure*}

The search loop combines grammar-based candidate generation, LLM
selection, adaptive parameter sampling, and deterministic physical
evaluation (Fig.~\ref{fig:workflow}). At each iteration, the LLM receives
a finite set of legal candidates together with a summary of the current
search history and returns a candidate index and rationale. The selected
structure is compiled into a Lagrangian and paired with parameter samples
generated by the numerical sampler. LangGraph coordinates these steps
\cite{LangGraph2024}; the validation harness runs the EOM, Builder, and
Solver stages and records their outcomes. The LLM's role is to choose
where the pipeline searches next.

The search strategy adapts using only current-run memory.  The memory contains validation-stage reachability, failure tags, accepted and rejected sample rows, $\chi^2$ and information-criterion values, structural fingerprints, scalar-family statistics, and parameter priors inferred during the same experiment.  It does not contain historical Solver winners or best-fit vectors from earlier human or batch searches.  After each completed structural trial, the search code uses these facts to update three interpretable memory layers: a structure prior, a marginal parameter prior, and a structure-conditioned joint prior that associates successful parameter vectors with exact or nearby theory families. These priors guide subsequent proposals.

Each \emph{cold-start} run begins with the common grammar and prescribed
priors but an empty search history. This initialization excludes results
from previous searches; the LLM retains its pretrained weights and the
knowledge encoded in them. Its weights remain fixed throughout the run,
and adaptation takes place through the external proposal priors.

The search budget counts parent theory proposals separately from
parameter samples. Each parent is assigned 1000 initial parameter
samples. An accepted initial fit with $\chiT<2000$ triggers an
additional 3000-sample scan of the same structure; an expanded fit
with $\chiT<1638$ triggers a further 5000-sample scan. These additional
scans do not count as new parent proposals. Earlier symbolic or
numerical checks can reject a proposal before its sample allocation
is exhausted. The benchmark uses 1000 parents per LLM, while the
separate memory comparison uses 250 parents per setting.

\subsection{Proposal parameters and scan ranges}
\label{subsec:parameter-priors}

Table~\ref{tab:parameter-priors} summarizes the default cold-start ranges for the continuous component of the search.  They are exploration priors rather than physical acceptance conditions: a sample drawn from these ranges is still rejected if it fails the modified Friedmann constraint, violates the background EFT cut, produces an irregular trajectory, or admits no acceptable real root for the constrained initial velocity.

\begin{table*}[t]
\centering
\begin{ruledtabular}
\begin{tabular}{lllll}
Parameter & Dimensionless definition & Applies to & Default range & Sampling scale \\
\hline
$H_0$ & $H_0\,[{\rm km\,s^{-1}\,Mpc^{-1}}]$ & all sectors & $58$--$74$ & linear \\
$\Omega_{m0}$ & $\rho_{m0}/(3\Mp^2H_0^2)$ & all sectors & $0.22$--$0.40$ & linear \\
$\lambda_H$ & $\Lam/H_0$ & all sectors & $5.0\times10^2$--$8.0\times10^4$ & log \\
$\varphi_{1,0}$ & $\phi_1(t_0)/\Mp$ & $\phi_1$ sectors & $10^{-3}$--$10^{1}$ & log \\
$\mu_1$ & $m_1/H_0$ & $\phi_1$ sectors & $10^{-2}$--$10^{2}$ & log \\
$\bar f_1$ & $f_1/\Mp$ & $\phi_1$ axion sectors & $10^{-2}$--$10^{2}$ & log \\
$\varphi_{2,0}$ & $\phi_2(t_0)/\Lam$ & $\phi_2$ sectors & $10^{-5}$--$10^{5}$ in $|\varphi_{2,0}|$ & signed log \\
$q_{2,0}$ & $\dd\varphi_2/\dd\ln a|_{0}$ & two-scalar sectors & $10^{-5}$--$10^{5}$ in $|q_{2,0}|$ & signed log \\
$\mu_2$ & $m_2/\Lam$ & $\phi_2$ sectors & $10^{-5}$--$10^{5}$ & log \\
$\bar f_2$ & $f_2/\Lam$ & $\phi_2$ axion sectors & $10^{-5}$--$10^{5}$ & log \\
$c_I$ & non-EH operator coefficients & generated operators & $10^{-10}$--$10^{20}$ in $|c_I|$ & log, both signs \\
\end{tabular}
\end{ruledtabular}
\caption{Default continuous proposal ranges.  These are broad cold-start exploration priors, not physical acceptance cuts. In one-scalar models the initial scalar velocity is solved from the modified Friedmann constraint rather than independently sampled.  In two-scalar models $q_{2,0}$ is sampled and the constraint solves for the remaining scalar velocity.}
\label{tab:parameter-priors}
\end{table*}

\subsection{Likelihood and structure objectives}
\label{subsec:aic-objective}

For a fixed Lagrangian density $\mathscr{L}$, the parameter-search objective is
\begin{equation}
\chi^2_{\min}(\mathscr{L})
=
\min_{\bm{\theta}_{\mathscr{L}}}
\chi^2_{\rm tot}(\mathscr{L},\bm{\theta}_{\mathscr{L}}),
\label{eq:fixed-structure-chi2}
\end{equation}
where all parameter vectors $\bm{\theta}_{\mathscr{L}}$ have the same nominal dimension. In practice, reported minima are the lowest validated values found within the stated sampling budget. Raw $\chi^2_{\rm tot}$ is used for parameter-prior updates, sample ranking, and expansion or refinement decisions within that structure.

Different Lagrangians need not have the same number of likelihood-relevant free directions.  The principal cross-structure objective is consequently
\begin{equation}
\mathrm{AIC}(\mathscr{L})
=
\chi^2_{\min}(\mathscr{L})+2k(\mathscr{L}),
\label{eq:aic}
\end{equation}
where $k(\mathscr{L})$ is the nominal number of likelihood-relevant
free parameters after the derivative-order substitutions. The count includes parameters that remain free and enter the background likelihood, including the profiled supernova absolute-magnitude offset and BAO scale.
For the flat-$\Lambda$CDM reference used here,
\begin{equation}
k_{\Lambda{\rm CDM}}=3
\qquad
\mathrm{AIC}_{\Lambda{\rm CDM}}
=1645.28+2\times3=1651.28
\label{eq:lcdm-aic}
\end{equation}
corresponding nominally to $\Omega_{m0}$, the profiled supernova offset, and the profiled BAO scale.  We report
\begin{equation}
\Delta\mathrm{AIC}
=
\mathrm{AIC}(\mathscr{L})
-\mathrm{AIC}_{\Lambda{\rm CDM}}
\label{eq:delta-aic}
\end{equation}
for comparison under AIC. With this nominal parameter count, negative values favor the candidate under AIC, whereas positive values favor the reference.
This use of information criteria for cosmological model comparison
follows the discussion in Ref.~\cite{Liddle2007}.

We also evaluate the Bayesian information criterion (BIC) for the
same fixed-covariance Gaussian likelihood and nominal parameter count
\cite{Schwarz1978,Liddle2007},
\begin{align}
\mathrm{BIC}(\mathscr{L})
&=\chi^2_{\min}(\mathscr{L})+k(\mathscr{L})\ln N,
\label{eq:bic}\\
\Delta\mathrm{BIC}
&=\mathrm{BIC}(\mathscr{L})-\mathrm{BIC}_{\Lambda{\rm CDM}}.
\label{eq:delta-bic}
\end{align}
Here $N=1820+13=1833$ is the number of supernova and BAO data-vector
entries used in the likelihood, with the full within-dataset covariance
matrices retained. Using the same parameter count $k$ as for AIC and
the reported reference $\chi^2_{\Lambda{\rm CDM}}=1645.28$ gives
$\mathrm{BIC}_{\Lambda{\rm CDM}}=1645.28+3\ln(1833)\simeq1667.8211$.
All observed-data information-criterion differences in this paper,
including the supplementary controls, use these reported reference values.
Positive $\Delta\mathrm{BIC}$ favors flat $\Lambda$CDM under BIC.
AIC guides the cross-structure search by balancing background fit
against the nominal number of free parameters. A lower AIC ranks a
candidate more favorably under this criterion, but does not by itself
establish a better physical description than $\Lambda$CDM. We report
BIC as a complementary comparison and interpret both criteria within
the adopted likelihood and finite search budget. Neither criterion penalizes the structural search itself: both charge for the parameters of a candidate, not for the cost of having selected it from a large grammar-generated space. Description-length criteria, which price the specification of the selected model within the candidate set \cite{Desmond:2026MDL}, offer a complementary route to the trials-corrected comparison discussed in Sec. VI B and would be a natural extension of the present scoring.

\subsection{AIC-guided proposal updates}

The search code uses completed structural trials to update an
empirical proposal prior. LangGraph carries the candidate specifications,
validation diagnostics, and numerical results between iterations.

For each Lagrangian $\mathcal L_j$,
we define the score
\begin{equation}
S_j
=
{\rm AIC}_j-2k_{\Lambda{\rm CDM}}
=
\chi^2_{j,\min}
+2\left(k_j-k_{\Lambda{\rm CDM}}\right),
\label{eq:structure-score}
\end{equation}
where $k_{\Lambda{\rm CDM}}=3$.
The score is converted into a schematic reward,
\begin{equation}
r_j = g(S_j)-P_j^{\rm fail},
\label{eq:structure-reward}
\end{equation}
where $g$ decreases monotonically with $S_j$, while
$P_j^{\rm fail}$ incorporates diagnostic penalties from unsuccessful
EOM or Solver evaluations. The search code aggregates these rewards
over interpretable theory features, including scalar content,
potential type, operator count, and covariant operator content. The
resulting weights favor features associated with low-AIC accepted
fits and downweight features associated with systematic failures.

For the next iteration, these empirical weights reweight a fresh set
of grammar-legal candidates. The language-model agent then selects one
candidate from this finite, pre-ranked set, while a fixed exploration
floor preserves underexplored theory families.

\subsection{Coupled structure and parameter proposals}
\label{subsec:joint-learning}

Structure selection and parameter sampling are coupled in the search loop.
The proposal distribution is represented schematically as
\begin{equation}
p_t(\mathcal{L},\boldsymbol{\theta})
=
\pi_t(\mathcal{L})\,
q_t(\boldsymbol{\theta}\mid\mathcal{L}),
\label{eq:joint-proposal}
\end{equation}
where $\pi_t(\mathcal{L})$ is the adaptive proposal distribution over
Lagrangian structures, and $q_t(\boldsymbol{\theta}\mid\mathcal{L})$ is
the conditional proposal distribution for their parameters.

The coupling operates in both directions.  First, the operator and scalar
content of $\mathcal{L}$ determines which parameter experience is relevant:
the memory retrieves successful vectors from the exact theory family, a
coarser operator family, the same scalar family, or finally the global
fallback.  Second, the parameter scan determines the empirical quality of
the structure.  The best validated parameter vector fixes the AIC-based
reward assigned to that structural trial and therefore changes the
probability of related structures in later proposals.  Thus, a structure
is not rewarded independently of whether its viable parameter region can
actually be found.

The current-run memory consequently stores joint records
$(\mathcal{L},\boldsymbol{\theta},\chi_{\rm tot}^{2},
\mathrm{outcome})$. For a new structure, shared cosmological and scalar parameters
are seeded or perturbed using the nearest compatible joint records, while
coefficients belonging to its new operators are recomputed.  An explicit
exploration component continues to sample the marginal or default priors,
preventing the coupled memory from collapsing onto a single previously
successful family.

\section{Benchmark Results}
\label{sec:results}

\subsection{Observed-data search results}
\label{subsec:observed-search-results}

Every accepted background fit must pass the implemented physical and numerical
checks described above; the proposal agent cannot bypass this pipeline
(Fig.~\ref{fig:workflow}).
We benchmarked eight LLM candidate selectors using the same grammar,
deterministic physics pipeline, initial parameter priors, parameter-sampling
schedule, and eight CPU cores. Each run contains 1000 completed initial
theory proposals and maintains its own search history and proposal priors.
Claude Sonnet 5, Claude Opus 5, and Claude Fable 5 are evaluated in
separate runs within the Claude model family. The comparison concerns
their behavior as candidate selectors under a common search setup;
separate runs do not imply independent training histories for the
underlying pretrained models.

\begin{table*}[t]
\centering
\caption{
Matched cold-start benchmark for eight LLM candidate selectors.
The minimum $\chi_{\mathrm{tot}}^2$ and minimum AIC need not correspond
to the same theory. The nominal parameter count $k$ refers to the
minimum-AIC theory, with
$\Delta\mathrm{AIC}=\mathrm{AIC}-1651.28$.
The $\Delta\mathrm{BIC}$ column uses the same minimum-AIC
representative and $N=1833$, with
$\Delta\mathrm{BIC}=\mathrm{BIC}-[1645.28+3\ln(1833)]$.
The entries retain the nominal parameter counts used in the benchmark.
Here $N_{\mathrm{better}}$ denotes the number of parent initial theories
whose best AIC over the initial and any triggered expanded or refined
scan is below the flat-$\Lambda$CDM value. Acceptance is evaluated over
the 1000 completed initial proposals. The final column gives the mean
total API-token usage per initial theory proposal.
}
\label{tab:agent-benchmark}
\small
\setlength{\tabcolsep}{2.5pt}
\renewcommand{\arraystretch}{1.12}
\begin{ruledtabular}
\begin{tabular}{lrrrrrrrr}
Agent
& $\min\chi_{\mathrm{tot}}^2$
& $\mathrm{AIC}_{\min}$
& $\Delta\mathrm{AIC}$
& $\Delta\mathrm{BIC}$
& $k$
& $N_{\mathrm{better}}$
& Accepted (\%)
& Tokens/theory \\
\hline
GPT-5.6 Sol
& 1638.016 & 1648.016 & $-3.264$ & $+7.764$ & 5 & 130 & 23.9 & 5135 \\
Claude Sonnet 5
& 1637.807 & 1651.184 & $-0.096$ & $+16.445$ & 6 & 1 & 38.1 & 8006 \\
Claude Opus 5
& 1637.960 & 1647.961 & $-3.319$ & $+7.708$ & 5 & 135 & 35.4 & 7921 \\
Claude Fable 5
& 1637.958 & 1647.970 & $-3.310$ & $+7.717$ & 5 & 90 & 28.4 & 7502 \\
Gemini 3.6 Flash
& 1637.956 & 1647.956 & $-3.324$ & $+7.703$ & 5 & 197 & 36.3 & 6029 \\
Qwen 3.7 Max
& 1638.011 & 1648.011 & $-3.269$ & $+7.759$ & 5 & 176 & 29.7 & 7487 \\
DeepSeek V4 Pro
& 1637.961 & 1647.962 & $-3.318$ & $+7.709$ & 5 & 210 & 32.5 & 5645 \\
Kimi K3
& 1637.952 & 1647.952 & $-3.328$ & $+7.699$ & 5 & 160 & 32.7 & 5404 \\
\end{tabular}
\end{ruledtabular}
\end{table*}

\begin{figure*}[t]
\centering
\includegraphics[width=0.86\textwidth]{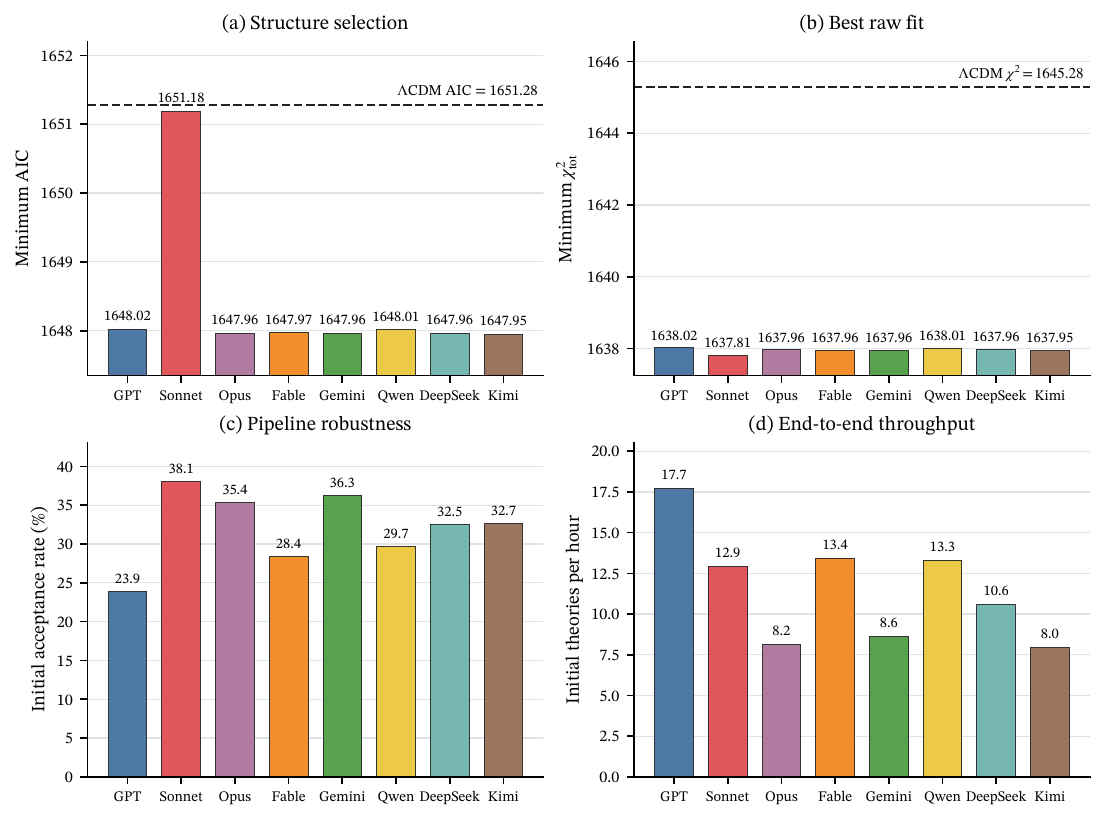}
\caption{Matched eight-agent benchmark. Panels (a) and (b) show the minimum AIC and minimum raw $\chi^2$; dashed lines are the corresponding flat-$\Lambda$CDM values. Panels (c) and (d) show initial-proposal acceptance and end-to-end throughput. Sonnet, Opus, and Fable denote separate search runs using the three Claude models. Numerical labels give the plotted values.}
\label{fig:benchmark-dashboard}
\end{figure*}

All eight search runs yield a raw fit below
$\chi^2_{\Lambda{\rm CDM}}=1645.28$ and at least one structure with
$\mathrm{AIC}<\mathrm{AIC}_{\Lambda{\rm CDM}}$.  Claude Sonnet 5
reaches the smallest raw value, $\chiT=1637.807$.  The lowest recorded
nominal AIC occurs in the Kimi K3 run, with
$\chiT=1637.952$, $\mathrm{AIC}=1647.952$, and
$\Delta\mathrm{AIC}=-3.328$. Seven agents reach a $k=5$ minimum-AIC
solution, and their minimum AIC values span only 0.065.
The finite-budget optimization scatter measured in
Appendix~\ref{sec:de-baseline} is comparable to this spread, so these
closely spaced minima do not establish a reliable ranking of the seven
agents based on minimum AIC alone.

\begin{figure}[t]
\centering
\includegraphics[width=\columnwidth]{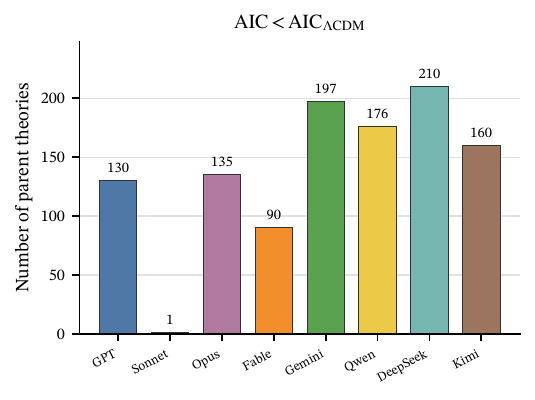}
\caption{Number of parent initial theories whose lowest AIC over the
initial and any triggered parameter scan is below the flat-$\Lambda$CDM
value.  Each parent theory is counted once.}
\label{fig:aic-improvement-counts}
\end{figure}

Across the 8000 parent proposals, 1099 achieve
$\Delta\mathrm{AIC}<0$ after their best evaluated parameter stage
(Fig.~\ref{fig:aic-improvement-counts}).  DeepSeek returns the largest
set, with 210 AIC-improved theories, followed by Gemini, Qwen, Kimi,
Opus, GPT, Fable, and Sonnet with 197, 176, 160, 135, 130, 90, and 1,
respectively. These counts measure how often the searches return
fits with lower AIC; they need not represent distinct physical theories,
as illustrated by the reductions in Sec.~\ref{sec:representatives}.
The proposals also share data and adaptive search histories, so these
counts cannot be treated as independent evidence against $\Lambda$CDM.

\begin{figure}[t]
\centering
\includegraphics[width=\columnwidth]{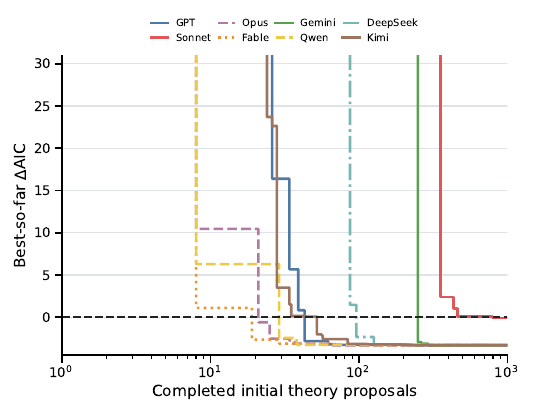}
\caption{Best-so-far $\Delta\mathrm{AIC}$ versus the parent
initial-theory number.  The dashed line is the flat-$\Lambda$CDM
baseline.}
\label{fig:aic-convergence}
\end{figure}

The competitive basin is usually reached early, although the first-hit
theory is model dependent (Fig.~\ref{fig:aic-convergence}).

\begin{figure}[t]
\centering
\includegraphics[width=\columnwidth]{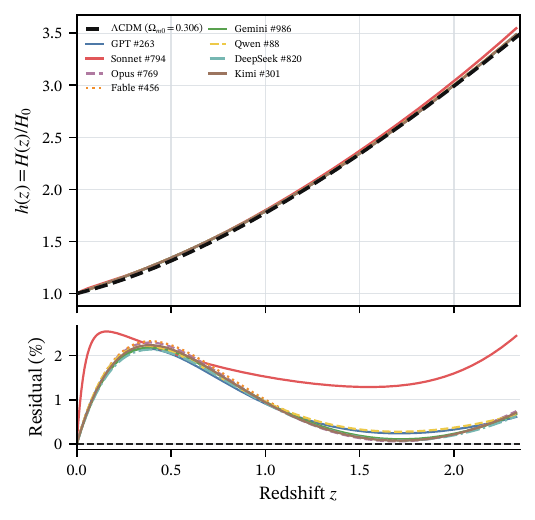}
\caption{Expansion histories for the minimum-AIC structure selected
from each agent.  The upper panel shows $h(z)=H(z)/H_0$ together with
flat $\Lambda$CDM; the lower panel gives the percentage residual from
the reference history.}
\label{fig:hubble-comparison}
\end{figure}

The runs differ in operational metrics: Sonnet has the highest
initial acceptance fraction, GPT the highest throughput, and DeepSeek
the most AIC-improved parent theories.  Despite these different search trajectories, the
minimum-AIC expansion histories remain close to flat $\Lambda$CDM over
the fitted redshift range (Fig.~\ref{fig:hubble-comparison}).  These fits give modest improvements in raw $\chi^2$ and nominal
AIC, while BIC favors flat $\Lambda$CDM for all eight reported
minimum-AIC representatives (Table~\ref{tab:agent-benchmark}).
We use these criteria to characterize the search outcomes, without
interpreting the AIC improvements as evidence for a departure from
$\Lambda$CDM.

For a phenomenological comparison, we also fitted the spatially flat
Chevallier--Polarski--Linder (CPL) parametrization,
$w(z)=w_0+w_a z/(1+z)$~\cite{ChevallierPolarski2001,Linder2003},
to the same 1820 DES-Dovekie SN and 13 DESI DR2 BAO entries, retaining
the full within-dataset covariance matrices. The SN offset and BAO scale
were analytically profiled as in Sec.~\ref{sec:observables}, without CMB
data or an external sound-horizon calibration. Assuming pressureless
matter and CPL dark energy, with radiation neglected as in the late-time
reference, numerical minimization gives
\begin{equation}
\begin{gathered}
\Omega_{m0}=0.31416,\\
w_0=-0.84344,\qquad w_a=-0.53571.
\end{gathered}
\label{eq:cpl-best-fit}
\end{equation}
These are best-fit values, without estimated parameter uncertainties.
Counting the three background parameters and two profiled nuisance
parameters gives $k=5$, $\chiT=1638.1199$, and
$\mathrm{AIC}=1648.1199$. Relative to the reported flat-$\Lambda$CDM
reference, $\Delta\mathrm{AIC}=-3.160$ and
$\Delta\mathrm{BIC}=+7.867$. The CPL AIC exceeds that of the recorded
quadratic-quintessence minimum, K301, by only 0.168; this small difference
indicates comparable background fit quality and does not establish a
meaningful preference between them.

The CPL best-fit parameters are close to the marginalized posterior
medians $(\Omega_{m0},w_0,w_a)=(0.313,-0.843,-0.53)$ reported for
DES-Dovekie plus DESI DR2 BAO in Table~10 of
Ref.~\cite{Popovic2026Dovekie}; these medians are distinct statistical
summaries from our best-fit point. CPL is used here solely as a
background parametrization: the fit specifies neither a covariant
action nor its perturbative stability.
Our profiled distance likelihood does not independently constrain $H_0$
and therefore does not test the quintessence--$H_0$ relation discussed in
Ref.~\cite{Colgain2026DESIEvolution}. Canonical quintessence cannot cross
$w=-1$, whereas such a crossing has been proposed as an interpretation of
differing SN, BAO and CMB matter-density constraints~\cite{Colgain2026FullShape}.
The comparable AIC values found here need not persist when additional
data are included.

\subsection{A \texorpdfstring{$\Lambda$CDM}{LambdaCDM} mock-data diagnostic}
\label{subsec:mock-lcdm250}

Searching many candidates can improve the fit by chance even when
$\Lambda$CDM generates the data. Interpreting a selected minimum therefore
requires accounting for the search procedure~\cite{GrossVitells2010}.
As a limited check under this null model, we searched 250 parent theories
on one realization generated from flat $\Lambda$CDM.

We retained the redshifts, observable definitions, and full likelihood
covariance matrices of the 1820 SN and 13 BAO entries used in the analysis.
The synthetic data vector was generated as
\begin{equation}
\begin{split}
\boldsymbol d^{\rm mock}
&=\boldsymbol m_{\Lambda{\rm CDM}}(\widehat{\boldsymbol\theta}_{\rm obs})
  +\boldsymbol\epsilon,\\
\boldsymbol\epsilon&\sim\mathcal N(\boldsymbol 0,C),\qquad
C=\operatorname{diag}(C_{\SN},C_{\BAO}).
\end{split}
\label{eq:mock-generation}
\end{equation}
The mean uses the observed-data $\Lambda$CDM fit with $\Omega_{m0}=0.306$,
its profiled SN offset and BAO scale, and the distance-scale convention
$H_0=70\,{\rm km\,s^{-1}\,Mpc^{-1}}$. The noise is a new Gaussian draw
with seed 20260908 and independent SN and BAO random streams; all supplied
correlations within each block are retained. For the SN precision matrix
$P_{\SN}=C_{\SN}^{-1}=LL^{\mathsf T}$, we generated the noise as
$L^{-\mathsf T}\boldsymbol z$ with standard-normal $\boldsymbol z$;
the BAO block uses the covariance Cholesky factor. No additional diagonal
noise is added. This fixed-covariance likelihood simulation does not
resimulate survey selection or covariance-estimation uncertainty.

We refitted $\Lambda$CDM to the same mock, profiling the SN offset and
BAO scale. The existing $\Omega_{m0}$ grid from 0.05 to 0.60 in steps of
0.001 gives $\Omega_{m0}=0.309$ and $\chi^2=1807.0318$, with
$k_{\Lambda{\rm CDM}}=3$. All mock AIC and BIC differences are measured
relative to this refitted reference, separately from the observed-data
comparisons.

The search used GPT-5.6 Sol, the 141-operator grammar, one to eight selected
operators per parent, and the scalaron-enabled pipeline described in
Appendix~\ref{app:scalaron-extension}. It retained the recorded settings
of the existing 250-parent observed-data run: open search, eight
asynchronous slots, search seed 20260607, and a 1200-s setup timeout.
Each parent received 1000 initial parameter samples; a finite accepted
fit with $\chi^2<2000$ triggered 3000 additional samples, and an expanded
fit with $\chi^2<1638$ triggered a further 5000 samples. These additional
scans do not count as new parents. The run completed all 250 parents with
initially empty search memory and isolated mock data and outputs.
The launch settings, then-current Oscar source, and grammar cache were
checked, subject to the historical-source limitation discussed in
Appendix~\ref{app:scalaron-extension}. Matching these settings does not
imply identical realized computational work: the absolute $\chi^2$ gates
can trigger different numbers of parameter evaluations on observed and
mock data.

\begin{table}[t]
\centering
\caption{Results of the search on the $\Lambda$CDM mock dataset. The reference and candidate
are fitted to the same single mock realization.  The candidate is the
lowest-AIC recorded fit after the 250-parent search (the expanded pass of
parent T172), using the nominal count $k$.  BIC follows the convention in
Eq.~\eqref{eq:bic}, with $N=1833$.}
\label{tab:mock-lcdm250}
\begin{ruledtabular}
\begin{tabular}{lrr}
Quantity & \shortstack{Mock $\Lambda$CDM\\refit} & \shortstack{Best mock\\candidate} \\
\hline
$\chi^2$             & 1807.0318 & 1805.7288 \\
$k$                  & 3         & 5         \\
$\mathrm{AIC}$       & 1813.0318 & 1815.7288 \\
$\Delta\mathrm{AIC}$ & 0         & $+2.6970$ \\
$\mathrm{BIC}$       & 1829.5729 & 1843.2974 \\
$\Delta\mathrm{BIC}$ & 0         & $+13.7245$ \\
\end{tabular}
\end{ruledtabular}
\end{table}

The best candidate lowers $\chi^2$ by 1.303 relative to the mock
$\Lambda$CDM refit, less than the AIC penalty for its two additional
nominal parameters:
\begin{equation}
\Delta\mathrm{AIC}^{\rm mock}
=\underbrace{-1.302965}_{\Delta\chi^2}
+\underbrace{4}_{2\Delta k}
=+2.697035.
\label{eq:mock-delta-aic}
\end{equation}
No recorded validated candidate in this run has
$\Delta\mathrm{AIC}^{\rm mock}<0$, and BIC also favors the refitted
mock $\Lambda$CDM reference (Table~\ref{tab:mock-lcdm250}).

One search on one noise realization cannot estimate a false-positive
rate or calibrate the look-elsewhere effect of the 8000-parent benchmark
across eight LLMs. Such a calibration would require independent mock
realizations with a search protocol and budget matched to the statistic
being assessed. We therefore assign no global $p$-value or
trials-corrected significance to the observed-data AIC minima.

\subsection{Search diagnostics and controls}
\label{subsec:search-controls}

The acceptance fractions in Table~\ref{tab:agent-benchmark} measure
whether the initial 1000-sample evaluation returns at least one accepted
background fit; they do not require an improvement over $\Lambda$CDM.
Across the eight runs, 2570 of 8000 initial proposals meet this criterion
(32.1\%). Most unsuccessful evaluations return no accepted fit at the
Solver stage. The recorded reasons are dominated by constraint drift
and failures to obtain acceptable initial scalar velocities, with
symbolic preparation and numerical integration also contributing.
Appendix~\ref{app:acceptance-diagnostics} gives the stage counts and
rejection diagnostics. These outcomes describe the search under its
chosen priors and computational limits, rather than a physical exclusion
of every unsuccessful theory.

To provide a reference for the guided searches, we also evaluated 1000
random theory proposals using the same observed data and 141-operator
grammar, with fixed proposal priors and no LLM selection or learned
feedback. Of these, 103 yielded an accepted initial fit (10.3\%).
The best result over the initial and triggered additional scans has
$\chiT=1645.214$ and $\mathrm{AIC}=1655.214$, corresponding to
$\Delta\mathrm{AIC}=+3.934$. Thus, this random search reaches a
background fit close to the $\Lambda$CDM reference, but none of its
evaluated candidates improves on the reference AIC. Its acceptance
fraction and best AIC are less favorable than those of the eight
recorded LLM-guided runs. The sampling procedure and scope of this
comparison are described in Appendix~\ref{app:random-search}.

In a separate comparison using the scalaron-enabled pipeline,
we evaluated GPT-5.6 Sol with and without current-run memory, using
250 parent proposals per setting. The memory-enabled reference is
the run described in Appendix~\ref{app:scalaron-extension}, rather
than a prefix of the 1000-parent benchmark. The best AIC was 1648.035
with memory and 1652.025 without it, a difference of 3.990. The
memory-enabled run thus reached a lower minimum AIC, while the
initial acceptance fractions were similar (38.4\% and 39.6\%,
respectively). Appendix~\ref{app:memory-ablation} gives the settings
and limitations of this single-run comparison.

After the derivative-order constraints are imposed, the minimum-AIC
representatives from seven of the eight searches reduce to the same canonical
quintessence basin with a quadratic scalar potential in
Table~\ref{tab:reduced-basin}. Claude Sonnet 5 instead selects an
axion-like potential under the original nominal parameter count.
The following section examines these reduced structures and their
background dynamics.

\section{Representative Structures}
\label{sec:representatives}

We now examine representative structures returned by the search and
their late-time background dynamics. Quoted parameters specify
best-fit points found within the stated numerical budgets. We have not computed observational confidence or
credible intervals for these parameters. The reported digits and the AIC
dispersions in Table~\ref{tab:reduced-basin} and
Appendix~\ref{sec:de-baseline} therefore do not represent parameter
uncertainties.

The complete action used in the background calculation is
\begin{equation}
\calS_{\rm tot}=
\int d^4x\,\sqrt{-g}(\mathcal{L}+\mathcal{L}_{m})
\end{equation}
where $\mathcal{L}_{m}$ denotes a minimally coupled pressureless-matter
sector satisfying
$\rho_m(a)=\rho_{m0}a^{-3}$.
Because this matter sector is fixed and common to every candidate,
the representative expressions below give $\Lag$, the searched
gravity--dark-energy Lagrangian, without the measure factor $\sqrt{-g}$.

\subsection{Quintessence Models}

The minimum-AIC representative from the Kimi run, K301, reduces to
\begin{equation}
\Lag_{1}
= \left[
\frac{\Mp^2}{2}R-\frac12\nabla_\mu\phi_1\nabla^\mu\phi_1-\frac12m_1^2\phi_1^2
\right].
\label{eq:quadratic-basin}
\end{equation}
This reduced Lagrangian belongs to the canonical quadratic-quintessence
class \cite{RatraPeebles1988,CaldwellLinder2005}.

It gives
\(\chiT=1637.9518\), \({\rm AIC}=1647.9518\),
\({\rm BIC}=1675.5203\), and \(k=5\).
Relative to flat $\Lambda$CDM, this gives $\Delta\mathrm{AIC}=-3.328$
and $\Delta\mathrm{BIC}=+7.699$: AIC modestly favors this candidate,
whereas BIC favors the reference. The fitted scalar mass is $1.465\times10^{-33}\,\mathrm{eV}$, corresponding to $m_1/H_0=1.006$.

The recovery of quintessence illustrates how symbolic reduction and
AIC-based selection organize the prescribed theory space. The searches
include higher-curvature and scalar--gravity operators, yet repeatedly
return the familiar limit of general relativity coupled to a single
canonical massive scalar.

The fitted parameters determine the scalar energy budget and equation
of state. The present field amplitude is
\[
\phi_{1,0}=1.9811\,\Mp ,
\]
which, together with \(m_1/H_0=1.006\), gives
\[
\Omega_{V,0}
\equiv
\frac{V(\phi_{1,0})}{3\Mp^2H_0^2}
=
\frac{1}{6}
\left(\frac{m_1}{H_0}\right)^2
\left(\frac{\phi_{1,0}}{\Mp}\right)^2
\simeq 0.6620 .
\]
The present scalar velocity is
\[
q_{1,0}\equiv
\left.\frac{d\varphi_1}{d\ln a}\right|_0=-0.6297
\]
The initial velocity gives
\[
\Omega_{K,0}=\frac{q_{1,0}^2}{6}=0.0661,
\]
and hence
\[
\Omega_{\phi,0}=0.7281,
\qquad
w_{\phi,0}=-0.818 .
\]

The reconstructed evolution is nonmonotonic over the fitted interval:
\(w_\phi\) changes from approximately \(-0.08\) at \(z=2.33\) to
\(-0.85\) near \(z=0.38\), before reaching \(-0.82\) today. Moreover,
the scalar fraction remains appreciable at the upper end of the data,
with \(\Omega_\phi(z=2.33)\simeq0.175\). Thus, $\Lag_1$ is not simply a
numerically degenerate representation of a cosmological constant:
its close background fit is produced by a dynamically evolving scalar
whose high-redshift contribution provides a concrete target for
subsequent CMB and structure-growth tests.

\begin{table*}[t]
\label{quintessence models}
\centering
\caption{Minimum-AIC representatives in the quadratic-quintessence basin.
All listed representatives have $k=5$; $\Delta\mathrm{AIC}$ and
$\Delta\mathrm{BIC}$ use the same flat-$\Lambda$CDM reference as
Table~\ref{tab:agent-benchmark}.
For each agent, $N_{\rm basin}$ includes all similar parent theories,
each represented by its lowest AIC across all scan stages, and
$\sigma_{\rm basin}$ is the sample standard deviation of those per-parent
minima, including parents with poor fits. It is not an uncertainty on the
best-fit parameters or a measure of optimizer scatter.}
\label{tab:reduced-basin}
\small
\setlength{\tabcolsep}{2.5pt}
\begin{ruledtabular}
\begin{tabular}{lcrrrcccc}
Agent & Proposal & AIC & $\Delta\mathrm{AIC}$ & $\Delta\mathrm{BIC}$ & $\phi_{1,0}/M_p$ & $m_1\ [{\rm GeV}]$ & $N_{\rm basin}$ & $\sigma_{\rm basin}$ \\
\hline
GPT-5.6 Sol      & G263  & 1648.0163 & $-3.264$ & $+7.764$ & 1.783 & $1.714\times10^{-42}$ & 84  & 0.0592 \\
Claude Opus 5    & O769  & 1647.9609 & $-3.319$ & $+7.708$ & 1.952 & $1.349\times10^{-42}$ & 111 & 3.6327 \\
Claude Fable 5   & F456  & 1647.9700 & $-3.310$ & $+7.717$ & 1.947 & $1.429\times10^{-42}$ & 50  & 6.1689 \\
Gemini 3.6 Flash & Gm986 & 1647.9559 & $-3.324$ & $+7.703$ & 2.014 & $1.499\times10^{-42}$ & 159 & 711.6485 \\
Qwen 3.7 Max     & Q88   & 1648.0111 & $-3.269$ & $+7.759$ & 1.761 & $1.519\times10^{-42}$ & 137 & 1.4703 \\
DeepSeek V4 Pro  & D820  & 1647.9619 & $-3.318$ & $+7.709$ & 2.032 & $1.257\times10^{-42}$ & 210 & 145.1785 \\
Kimi K3          & K301  & 1647.9518 & $-3.328$ & $+7.699$ & 1.981 & $1.465\times10^{-42}$ & 118 & 2.4840 \\
\end{tabular}
\end{ruledtabular}
\end{table*}

Under the original nominal parameter count, Claude Sonnet's minimum-AIC
representative, C794, reduces to a similar structure with a different potential
\begin{align}
\Lag_2
= \Bigg\{&
\frac{\Mp^2}{2}R-\frac12\nabla_\mu\phi_1\nabla^\mu\phi_1
\nonumber\\[-2pt]
&-m_1^2f_1^2
\left[1-\cos\left(\frac{\phi_1}{f_1}\right)\right]
\Bigg\}.
\label{eq:claude-min-aic}
\end{align}
The periodic potential is the standard pNGB/axion-like dark-energy
form \cite{Frieman1995}.
It has \(\chi^2_{\rm tot}=1639.1838\),
\({\rm AIC}=1651.1838\), \({\rm BIC}=1684.2660\), \(k=6\), and
\begin{equation}
m_1=2.135\times10^{-41}\ {\rm GeV}.
\end{equation}

The distinctive behavior of $\Lag_{2}$ is not simply its periodic potential,
but a dynamically generated hilltop turnaround. Its fitted parameters,
\[
\frac{m_1}{H_0}=14.046,\qquad
\frac{f_1}{M_p}=0.07544,
\]
set the potential-barrier height to
\[
\frac{V_{\max}}{3M_p^2H_0^2}
=\frac{2}{3}
\left(\frac{m_1}{H_0}\right)^2
\left(\frac{f_1}{M_p}\right)^2
\simeq0.7485 .
\]
At \(z=2.33\), the field has
\(\phi_1/f_1\simeq1.69\) and \(q_1>0\), and therefore moves toward
the maximum of the cosine potential. It reaches
\(\phi_1/f_1\simeq3.128\simeq\pi\) near \(z\simeq0.66\), where
\(q_1\simeq0\) and \(w_\phi\simeq-1\). The scalar then reverses
direction and rolls away from the hilltop. Today,
\[
\frac{\phi_{1,0}}{f_1}=1.844,\qquad
q_{1,0}=-1.132,\qquad
w_{\phi,0}=-0.380 .
\]
$\Lag_{2}$ therefore realizes transient rather than asymptotic acceleration:
the reconstructed background accelerates approximately over
\(0.009\lesssim z\lesssim0.69\) and has already returned to mild
deceleration at \(z=0\). Its competitive distance fit is obtained
because the scalar closely mimics a cosmological constant during the
intermediate-redshift interval to which the integrated SN and BAO
distances are most sensitive.

With $\Lag_{2}$ as the axion-hilltop exception, the minimum-AIC
representatives returned by the other seven agents belong to the same
canonical quadratic-quintessence family as K301 after the
derivative-order constraints are imposed. Their fitted
scalar masses and present-day field amplitudes are reported in
Table~\ref{tab:reduced-basin}. Despite originating from different
proposal agents, these solutions consistently select a scalar mass of
order the present Hubble scale and a present-day field displacement of
order the reduced Planck mass. The separate search runs thus return closely similar parameter
regimes as well as the same reduced theory structure.

\subsection{Equivalent Axion-Quintessence Representation}

Claude Sonnet 5's smallest $\chi^2$ fit, C962, contains one additional surviving
operator:
\begin{align}
\Lag_3
= \Bigg\{&
\frac{\Mp^2}{2}R+\vartheta\phi_1\Box\phi_1-\frac12\nabla_\mu\phi_1\nabla^\mu\phi_1
\nonumber\\[-2pt]
&-m_1^2f_1^2
\left[1-\cos\left(\frac{\phi_1}{f_1}\right)\right]
\Bigg\},
\label{eq:claude-best-chi2}
\end{align}
where \(\vartheta=10^{-4}\).
Using the original
nominal parameter count, the search reported
\(\chi^2_{\rm tot}=1637.8067\), \({\rm AIC}=1651.8067\),
\({\rm BIC}=1690.4027\),
\(k=7\), and \(m_1=2.135\times10^{-41}\ {\rm GeV}\).
This nominal AIC ranking changes once the redundant kinetic normalization
is removed, as discussed below.

This nominal parameter count overestimates the number of physically
independent directions. Since \(\vartheta\) is constant, integration by
parts gives, up to a boundary term,
\[
\vartheta\phi_1\Box\phi_1
\doteq
-\vartheta\nabla_\mu\phi_1\nabla^\mu\phi_1 ,
\]
so that the kinetic coefficient becomes \(1+2\vartheta\). The invertible
field redefinition
\[
\begin{aligned}
\widetilde{\phi}_1&=\sqrt{1+2\vartheta}\,\phi_1,\\
\widetilde f_1&=\sqrt{1+2\vartheta}\,f_1,\\
\widetilde m_1&=\frac{m_1}{\sqrt{1+2\vartheta}},
\end{aligned}
\]
returns the action to the canonical axion form. Therefore
\(\vartheta\) is not an independent physical parameter, and the effective
count is \(k_{\rm eff}=6\), giving
\({\rm AIC}_{\rm eff}=1649.8067\) and
\({\rm BIC}_{\rm eff}=1682.8890\). C962 and C794 consequently belong
to the same theory equivalence class. With the corrected count, C962 has
a lower AIC than C794; Table~\ref{tab:agent-benchmark} retains the
original nominal counts and selected representatives. Their appearance as
distinct proposals exposes a failure mode of the present search: the agent can
identify a redundant representation without recognizing its simpler
field-redefined form. Future implementations can address this by
canonicalizing proposals under integration by parts and invertible field
redefinitions, supplemented by a numerical Hessian-rank test before
assigning the effective parameter count.

\section{Non-Einstein Hilbert Structures}
\label{sec:non-eh}

To explore structures without a bare Einstein--Hilbert term, we
performed a separate search that excluded this operator at proposal
time. Its different proposal restriction distinguishes it from the
benchmark in Sec.~\ref{sec:results}. The selected examples below have
$\chiT<2000$ but fit the data less well than the benchmark minima.
They illustrate the structures reached by this search and the
additional checks needed to assess their viability.

The minimum-AIC representatives in the preceding section reduce to
single-field models involving $\phi_1$. Here, $\phi_2$ enters
nonminimal curvature couplings and can set the effective gravitational
strength. We examine this role in an induced-gravity example and an
FLRW-equivalent scalar--Gauss--Bonnet representation.

\subsection{Induced-gravity structure}
\label{sec:induced-gravity-branch}

The DeepSeek V4 Pro search identified a non--Einstein--Hilbert
two-field structure,
\begin{align}
\label{eq:induced-L3}
\mathcal L_4 ={}&
\frac{1}{2}M_*^2(\phi_2)R
+\phi_1F(\phi_2)\Box\phi_1\\
&-\frac{1}{2}\nabla_\mu\phi_1\nabla^\mu\phi_1
-\frac{1}{2}\nabla_\mu\phi_2\nabla^\mu\phi_2
-V(\phi_1,\phi_2),
\nonumber\\
M_*^2(\phi_2)\equiv{}&
2\alpha M_p^2\frac{\phi_2^2}{\Lambda^2},
\qquad
F(\phi_2)\equiv
\frac{\beta}{\Lambda}\phi_2
+\frac{\gamma}{\Lambda^2}\phi_2^2,
\nonumber\\
V(\phi_1,\phi_2)\equiv{}&
m_1^2f_1^2
\left[
1-\cos\left(\frac{\phi_1}{f_1}\right)
\right]
+\frac{1}{2}m_2^2\phi_2^2 .
\end{align}

Equation~\eqref{eq:induced-L3} contains no field-independent
Einstein--Hilbert term. Instead, the background value of $\phi_2$
dynamically determines the effective Planck mass $M_*^2(\phi_2)$.
The normalization of the gravitational sector is therefore set by
a cosmological scalar background rather than by a fixed coefficient
of $R$, which is the defining induced-gravity feature
\cite{Zee1979,Zee1981}. 

The derivative interaction also has a transparent covariant
interpretation. Integration by parts gives
\begin{equation}
\phi_1F(\phi_2)\Box\phi_1
\doteq
-F(\phi_2)
 \nabla_\mu\phi_1\nabla^\mu\phi_1
-\phi_1F_{,\phi_2}
 \nabla_\mu\phi_1\nabla^\mu\phi_2,
\label{eq:induced-ibp}
\end{equation}
where $\doteq$ denotes equality up to a covariant boundary term.
The scalar sector may consequently be written as
\begin{equation}
-\frac{1}{2}\mathcal K_{IJ}
\nabla_\mu\phi^I\nabla^\mu\phi^J,
\qquad
\mathcal K_{IJ}
=
\begin{pmatrix}
1+2F(\phi_2) & \phi_1F_{,\phi_2}\\[3pt]
\phi_1F_{,\phi_2} & 1
\end{pmatrix},
\label{eq:induced-background-kinetic-matrix}
\end{equation}
with $\phi^I=(\phi_1,\phi_2)$. Thus, the same field $\phi_2$
that generates the effective gravitational coupling also controls
the normalization and kinetic mixing of the axion-like field
$\phi_1$.

With these definitions, the complete gravity--scalar Lagrangian
takes the compact form
\begin{equation}
\mathcal L_4
\doteq
\frac{1}{2}M_*^2(\phi_2)R
-\frac{1}{2}\mathcal K_{IJ}(\phi_1,\phi_2)
\nabla_\mu\phi^I\nabla^\mu\phi^J
-V(\phi_1,\phi_2).
\label{eq:induced-compact-lagrangian}
\end{equation}

On a spatially flat FLRW background, the corresponding modified
Friedmann equation can be written as
\begin{equation}
3M_*^2H^2
=
\rho_m
+\frac{1}{2}\mathcal K_{IJ}\dot{\phi}^I\dot{\phi}^J
+V
-3H\frac{\dd M_*^2}{\dd t}.
\label{eq:induced-Friedmann}
\end{equation}

The last term has no analogue in minimally coupled quintessence.
As $\phi_2$ evolves, it changes $M_*^2$ and hence the gravitational
normalization and expansion history. The curvature coupling, in turn,
enters the equation of motion for $\phi_2$.

The best trajectory on the
positive-$M_*^2$ branch gives
\begin{equation}
\chi^2_{\rm tot}=1718.247,
\qquad
\mathrm{AIC}=1746.247,
\qquad
k=14,
\label{eq:induced-rescan-statistics}
\end{equation}
with the principal couplings
\begin{align}
\alpha&=2.232\times10^6,\qquad\beta=9.551\times10^7,
\nonumber\\
\gamma&=4.427\times10^{-8},\nonumber\\
m_1&=2.818\times10^{-33}\text{eV},\qquad
m_2=1.572\times10^{-30}\text{eV},\nonumber\\
\phi_{1,0}&=2.525\times10^{18}\text{GeV},\nonumber\\
\phi_{2,0}&=-2.780\times10^{-33}\text{eV}.
\label{eq:induced-rescan-parameters}
\end{align}

Along the complete fitted trajectory,
\begin{equation}
1.000
\leq
\frac{M_*^2}{M_p^2}
\leq
2.088,
\qquad
c_T^2=1,
\qquad
0\leq z\leq2.33.
\label{eq:induced-rescan-Mstar}
\end{equation}
The effective Planck mass remains finite, positive, and of order
$M_p^2$ throughout the fitted interval. In particular, the trajectory
does not cross the singular surface $M_*^2=0$, and the coefficient of
the tensor kinetic term retains the correct sign. Together with
$c_T^2=1$, this establishes positive tensor kinetic and gradient
coefficients at quadratic order for the displayed action. The order-unity variation of $M_*^2$ further
shows that the $\phi_2^2R$ interaction is dynamically active, rather
than being equivalent to a constant renormalization of the
Einstein--Hilbert term.

These tensor-sector checks do not establish scalar stability.
Writing $A(\phi_2)\equiv M_*^2(\phi_2)$, the conformal transformation
$g^{\rm E}_{\mu\nu}=(A/M_p^2)g_{\mu\nu}$ gives the Einstein-frame
scalar kinetic matrix~\cite{Kaiser2010}
\begin{equation}
\mathcal K^{\rm E}_{IJ}
=\frac{M_p^2}{A}\mathcal K_{IJ}
+\frac{3M_p^2}{2A^2}A_{,I}A_{,J}.
\label{eq:induced-einstein-kinetic}
\end{equation}
Since $A$ depends only on $\phi_2$, its contribution from the conformal
transformation cannot change the $\phi_1$ diagonal entry,
$\mathcal K^{\rm E}_{11}=(M_p^2/A)[1+2F(\phi_2)]$.
At the quoted best-fit point,
$F(\phi_{2,0})\simeq-4.5210\times10^4$, so this coefficient
is already negative at $z=0$; it remains negative along the
evaluated fitted trajectory. This parameter point therefore
fails the scalar no-ghost condition despite its regular
background and positive tensor kinetic term. We retain it as
an example of a background fit rejected by a subsequent scalar-stability
check. This failure applies to the quoted fitted point and does not
exclude the entire induced-gravity theory family.

\subsection{Scalar--Gauss--Bonnet FLRW branch}
\label{sec:q809-branch}

After imposing the derivative-order relations
\(\beta=\zeta=0\) and \(\delta=-\epsilon\), the Qwen proposal Q809
reduces on a spatially flat FLRW background to
\begin{align}
\mathcal L_5
=
\Bigg[
&\frac{1}{2}F(\phi_2)R
+\xi(\phi_2)\mathcal G
\nonumber\\
&-\frac{1}{2}Z(\phi_2)
\nabla_\mu\phi_2\nabla^\mu\phi_2
-V(\phi_2)
\Bigg],
\label{eq:q809-flrw-theory}
\end{align}
where
\begin{equation}
\begin{aligned}
F(\phi_2)&=2\gamma\Mp^2\frac{\phi_2^2}{\Lam^2},
&
\xi(\phi_2)&=-\frac{\epsilon\Mp^2}{2\Lam^3}\phi_2,
\\
Z(\phi_2)&=1+2\eta-2\theta\frac{\phi_2^2}{\Lam^2},
&
V(\phi_2)&=\frac{1}{2}m_2^2\phi_2^2 .
\end{aligned}
\label{eq:q809-functions}
\end{equation}
The surviving coefficients are
\begin{equation}
\begin{gathered}
\gamma=-8.845\times10^3,
\qquad
\epsilon=4.123\times10^7,
\\
\eta=5.283\times10^{15},
\qquad
\theta=1.418\times10^{-10}.
\end{gathered}
\end{equation}
Thus, a single scalar controls a quadratic nonminimal coupling, a
linear scalar--Gauss--Bonnet interaction, and a noncanonical kinetic
normalization \cite{NojiriOdintsovSasaki2005}.  Since the action
contains no field-independent Einstein--Hilbert term, it is not a
small correction to general relativity.

The best Solver result is
\begin{equation}
\chi_{\rm tot}^2=1896.0009,
\qquad
{\rm AIC}=1918.0009,
\qquad
k=11,
\label{eq:q809-statistics}
\end{equation}
with
\begin{equation}
\begin{gathered}
H_0=61.262\ {\rm km\,s^{-1}\,Mpc^{-1}},
\qquad
\Omega_{m0}=0.32539,
\\
\frac{\Lam}{H_0}=1481.733,
\qquad
\frac{\phi_{2,0}}{\Lam}=-6.5997\times10^{-4},
\\
q_{2,0}=-6.4350\times10^{-3},
\qquad
\frac{m_2}{\Lam}=5.5063\times10^{-2},
\\
m_2=1.0662\times10^{-40}\ {\rm GeV},
\qquad
\frac{m_2}{H_0}=81.589 .
\end{gathered}
\label{eq:q809-initial-conditions}
\end{equation}
Relative to flat \(\Lambda\)CDM, this corresponds to
\(\Delta\chi^2=250.72\) and \(\Delta{\rm AIC}=266.72\).
Q809 is therefore retained for its unusual reduced structure rather
than for statistical competitiveness.

For the scalar--Gauss--Bonnet covariant completion of
Eq.~\eqref{eq:q809-flrw-theory}, the tensor coefficients and
propagation speed are~\cite{Kobayashi2011}
\begin{equation}
\mathcal G_T=F+8H\dot\xi,
\qquad
\mathcal F_T=F+8\ddot\xi,
\qquad
c_T^2=\frac{\mathcal F_T}{\mathcal G_T}.
\label{eq:q809-tensor-speed}
\end{equation}
Evaluating these quantities along the best-fit background at the
redshift of GW170817, \(z=0.009783\), gives
\begin{equation}
\frac{\mathcal G_T}{\Mp^2}=0.4618,
\qquad
\frac{\mathcal F_T}{\Mp^2}=1.7320,
\qquad
c_T^2=3.7508 .
\label{eq:q809-gw-result}
\end{equation}
The tensor kinetic and gradient coefficients are positive at this
redshift, but the low-energy propagation speed is substantially
nonluminal: $c_T-1\simeq0.937$ in units where the photon speed is unity.
If the displayed action remained applicable at LIGO frequencies,
such a deviation would be incompatible with GW170817
\cite{Abbott2017GW170817,CreminelliVernizzi2017,EzquiagaZumalacarregui2017}.
However, the fitted scale treated as the EFT cutoff here is
$\Lambda\simeq1.9363\times10^{-30}\,\mathrm{eV}$, far below the
$\sim10^{-13}\,\mathrm{eV}$ energies of LIGO-band waves. Extrapolating
the low-energy tensor speed to that band therefore requires information
about the theory above its cutoff~\cite{deRhamMelville2018}.
Equation~\eqref{eq:q809-gw-result} alone does not establish an
observational exclusion of this completion.

The original curvature-squared covariant parent also differs from
this Gauss--Bonnet completion by a Weyl-squared interaction,
\begin{equation}
\mathcal L_{\rm parent}-\mathcal L_5
\doteq\frac{3\epsilon M_p^2}{2\Lambda^3}\phi_2
C_{\mu\nu\rho\sigma}C^{\mu\nu\rho\sigma},
\label{eq:q809-weyl-remainder}
\end{equation}
where $C_{\mu\nu\rho\sigma}$ is the Weyl tensor. This follows from
the four-dimensional curvature identities after imposing the
coefficient relations above. The additional term vanishes on FLRW
but contributes to quadratic perturbations. Thus the tensor
coefficients in Eq.~\eqref{eq:q809-tensor-speed} apply to the
specified Gauss--Bonnet completion; the original parent requires
a separate perturbation analysis.

\section{Conclusions}

\label{sec:conclusion}

We have presented an automated framework for searching a prescribed
space of covariant modified-gravity models against cosmological data.
It combines grammar-based candidate generation, LLM selection, and
adaptive numerical parameter sampling with a deterministic physics
pipeline. The latter derives the FLRW dynamics, applies the implemented
background and numerical checks, and evaluates the DES-SN and DESI-DR2
BAO likelihood.

Seven of the eight separately initialized searches recovered the
quadratic-quintessence basin without using previous best-fit results as
search seeds. All eight returned at least one candidate with a nominal
AIC below the flat-$\Lambda$CDM reference. For the lowest-AIC candidate,
$\Delta\mathrm{AIC}=-3.328$ and $\Delta\mathrm{BIC}=+7.699$;
BIC favors $\Lambda$CDM for all eight reported minimum-AIC
representatives. The modest AIC gains do not establish an observational
preference for these models over flat $\Lambda$CDM. They show that the
search can recover familiar scalar-field models from the broader grammar
and locate parameter regions with competitive background fits.

The null-model diagnostic in Sec.~\ref{subsec:mock-lcdm250} applies the
search to one $\Lambda$CDM mock realization. After 250 parent proposals,
the lowest candidate AIC remains above the refitted mock reference by
2.697. This provides a limited check of the search on synthetic data;
it does not determine the global significance of the observed-data
minima. A calibration of the full search would require repeated mock
realizations with the corresponding search protocol and budget.

The additional controls provide a reference for the search procedure.
A 1000-parent fixed-prior random search reached a higher minimum AIC
and a lower initial acceptance fraction than the recorded LLM-guided
runs. In a separate 250-parent GPT-5.6 Sol comparison, the run with
current-run memory reached a minimum AIC lower by 3.990 than the run
without it, while their initial acceptance fractions were similar.
These single-run comparisons support further study of guided and
adaptive search; repeated runs are needed to measure the variability
of the differences.

The exploratory non-Einstein--Hilbert examples also show why background
fitting must be followed by additional physical checks. The quoted
induced-gravity fit fails the scalar no-ghost condition. The
evaluated scalar--Gauss--Bonnet completion has a substantially
nonluminal low-energy tensor speed, although applying the GW170817
constraint requires establishing the theory's validity at the observed
frequencies. Its original covariant parent shares the FLRW background
equations but differs at the perturbative level through a Weyl-squared
interaction.

The framework provides a systematic way to evaluate models beyond
small manually selected collections while retaining both successful
fits and failure diagnostics. Extending the validation to cosmological
perturbations, scalar and tensor stability, and structure-growth,
weak-lensing, CMB, redshift-space-distortion, and gravitational-wave
observations would allow future searches to assess viability beyond
the homogeneous background. A complementary direction is model selection by description length, which prices the search over structures rather than only the parameters within one.

\section*{acknowledgments}
We acknowledge useful conversations with Lawrence Edmond IV, Tucker Manton, Humberto Gilmer, and Heliudson Bernardo. C.N. thanks Tucker Manton for suggesting the minisuperspace approach used in the calculation. S.A. acknowledges support from the Simons Foundation, Award 896696. L.G. is supported by the DOE, Office of Science, Office of High Energy Physics under Award No. DE-SC0010010 and the Early Career Research program under Award No. DE-SC0026288. S.M.K. is supported by National Science Foundation (NSF) Grant No. PHY-2412666. We acknowledge support from the Brown University Division of Research through seed grant GR300480.

\appendix

\section{Differential-evolution parameter baseline}
\label{sec:de-baseline}

\begin{table*}[t]
\centering
\caption{Agent results compared with the differential-evolution
parameter-optimization baseline. The DE mean and sample standard deviation
are calculated from the minimum AIC values recovered in 20 independent
5000-evaluation restarts of each fixed Lagrangian.}
\label{tab:de-baseline}
\small
\setlength{\tabcolsep}{10pt}
\begin{ruledtabular}
\begin{tabular}{lcrr}
Agent
& Proposal
& Agent AIC
& $\langle{\rm AIC}_{\rm DE}\rangle\pm\sigma_{\rm DE}$ \\
\hline
GPT-5.6 Sol      & G263  & 1648.0163 & $1648.0147\pm0.1277$ \\
Claude Opus 5    & O769  & 1647.9609 & $1647.9825\pm0.0315$ \\
Claude Fable 5   & F456  & 1647.9700 & $1647.9934\pm0.0299$ \\
Gemini 3.6 Flash & Gm986 & 1647.9559 & $1647.9982\pm0.0564$ \\
Qwen 3.7 Max     & Q88   & 1648.0111 & $1647.9838\pm0.0345$ \\
DeepSeek V4 Pro  & D820  & 1647.9619 & $1647.9815\pm0.0308$ \\
Kimi K3          & K301  & 1647.9518 & $1647.9844\pm0.0329$ \\
\end{tabular}
\end{ruledtabular}
\end{table*}

The dispersion reported in Table~\ref{tab:reduced-basin} is calculated
across structurally similar but nonidentical parent proposals. To measure
the variability of the parameter search for a fixed structure, we
independently re-optimized the continuous parameters of each of the seven
representative Lagrangians using differential evolution
(DE)~\cite{StornPrice1997}.

We used the \texttt{DE/best/1/bin} strategy, which generates trial
parameters from the current best solution and population differences,
applies binomial crossover, and retains the solution with the lower AIC.
Each run started from the same broad parameter bounds used at the start
of the agent searches, without agent memory or historical best-fit seeds.
We performed 20 independent restarts for each Lagrangian, with 5000
Solver evaluations per restart. Since the structure and parameter count
were fixed, minimizing AIC was equivalent to minimizing $\chi^2_{\rm tot}$.
The standard deviation $\sigma_{\rm DE}$ describes the spread of the
best-fit AIC across these restarts, rather than an observational
parameter uncertainty.

Table~\ref{tab:de-baseline} summarizes the results. For six of the seven
representatives, $\sigma_{\rm DE}<0.06$; the largest value is $0.13$.
Every best-fit AIC obtained by the agents lies within one
$\sigma_{\rm DE}$ of the corresponding DE mean, and five of the seven
agent values lie below that mean. These results support the numerical
reproducibility of the fixed-structure fits under the tested settings.
The agent searches and DE restarts used different evaluation budgets,
so this comparison does not establish their relative optimization
efficiency.



\section{Individual Agent Search Trajectories}
\label{sec:agent-trajectories}

This appendix shows the search trajectory of each agent separately. For
every completed parent theory, the curves report the running-best
$\Delta{\rm AIC}$ relative to flat $\Lambda$CDM and terminate when the
minimum-AIC model of that agent is first reached. The horizontal axis is
linear, while the vertical axis uses a symmetric logarithmic scale with a
linear interval between $-10$ and $10$. Marker shapes denote the evaluation
stage: circles, squares, and stars indicate
the initial 1000-sample evaluation, the expanded 3000-sample evaluation,
and the 5000-sample refinement, respectively. The annotation ``final'' marks
the minimum-AIC point reached in each run, regardless of its evaluation stage.
The annotations identify
the proposal number and the structural or parameter-level change associated
with each improvement.

\begin{figure*}[p]
\centering
\includegraphics[width=0.96\textwidth,height=0.42\textheight,keepaspectratio]
{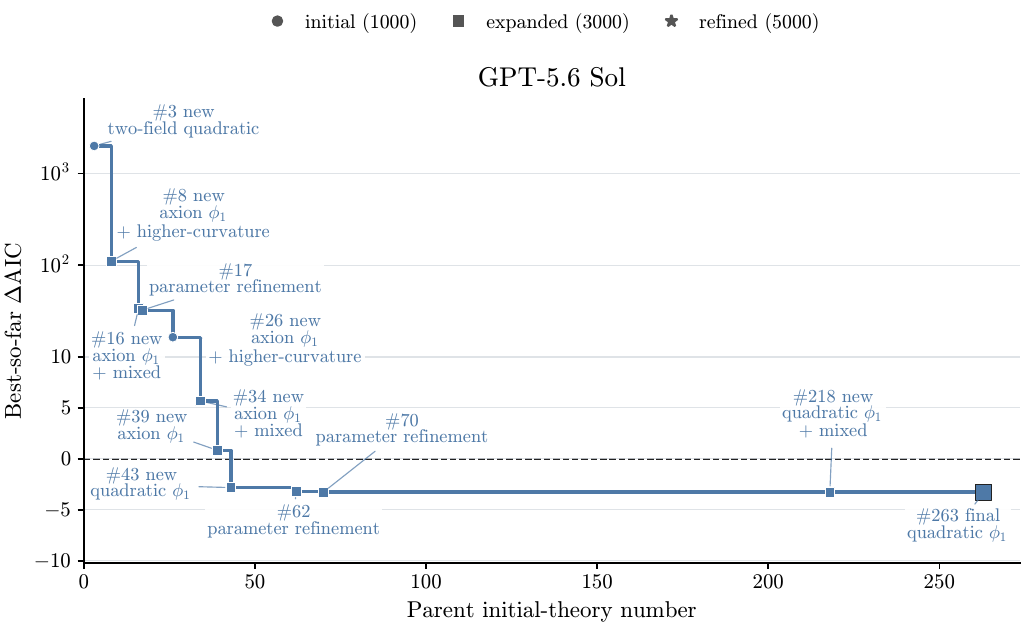}
\caption{Search trajectory of GPT-5.6 Sol.}
\label{fig:trajectory-gpt}
\par\medskip
\centering
\includegraphics[width=0.96\textwidth,height=0.42\textheight,keepaspectratio]
{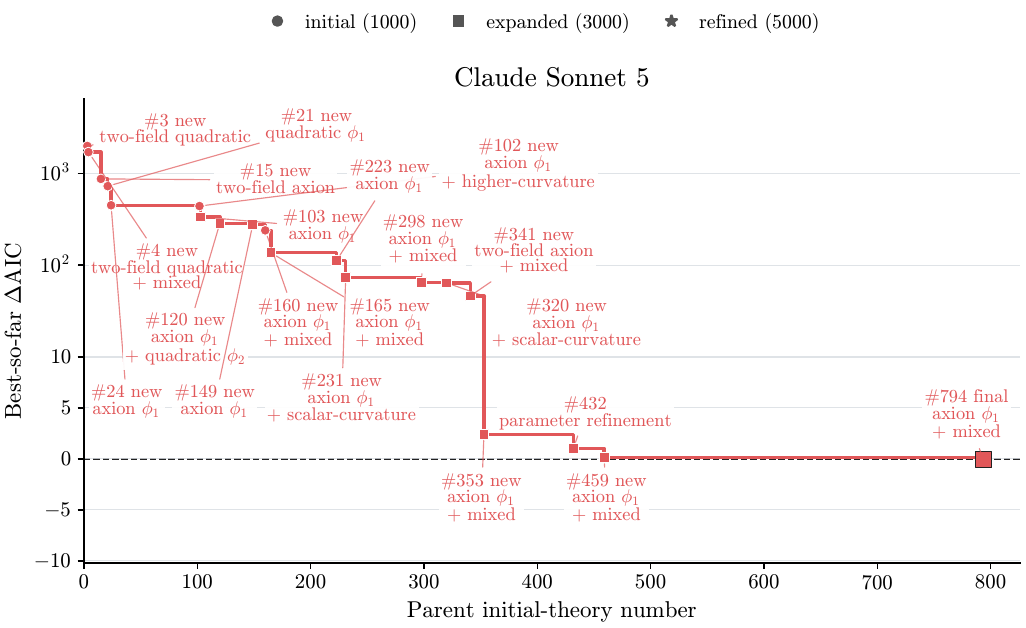}
\caption{Search trajectory of Claude Sonnet 5.}
\label{fig:trajectory-sonnet}
\end{figure*}

\begin{figure*}[p]
\centering
\includegraphics[width=0.96\textwidth,height=0.42\textheight,keepaspectratio]
{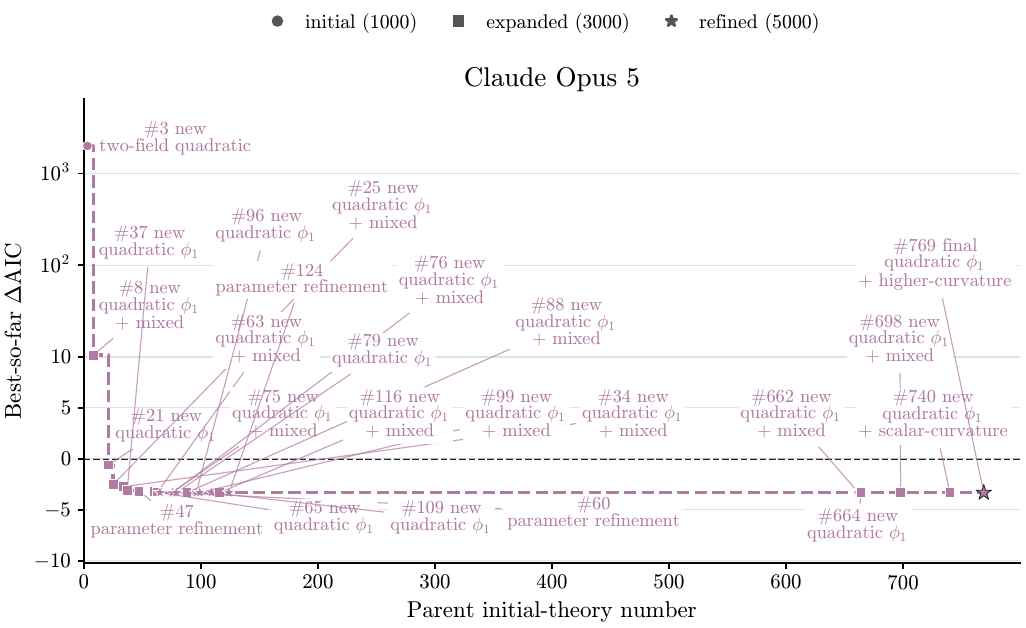}
\caption{Search trajectory of Claude Opus 5.}
\label{fig:trajectory-opus}
\par\medskip
\centering
\includegraphics[width=0.96\textwidth,height=0.42\textheight,keepaspectratio]
{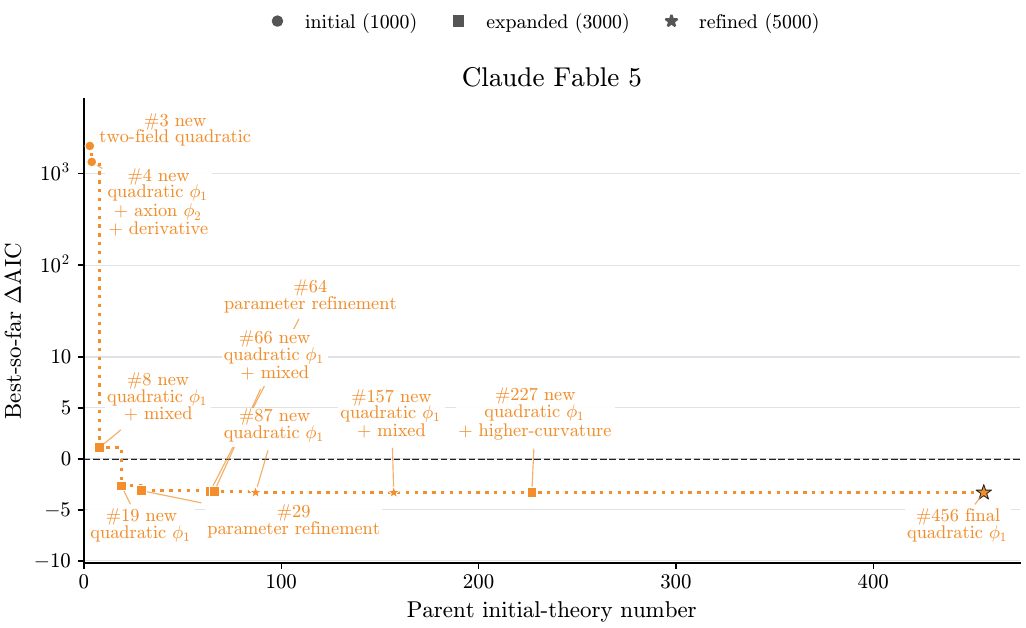}
\caption{Search trajectory of Claude Fable 5.}
\label{fig:trajectory-fable}
\end{figure*}

\begin{figure*}[p]
\centering
\includegraphics[width=0.96\textwidth,height=0.42\textheight,keepaspectratio]
{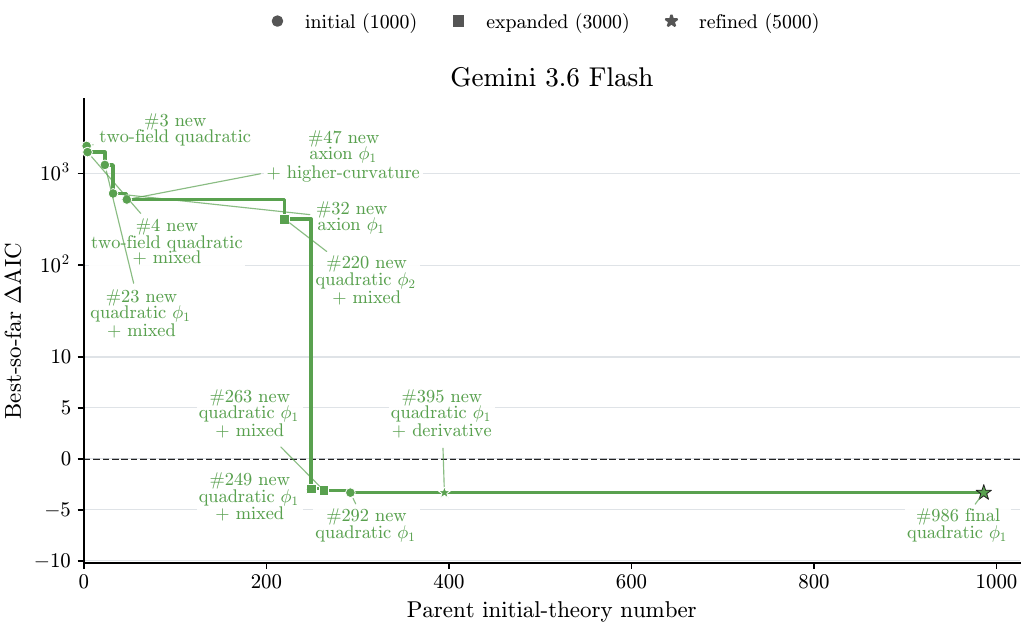}
\caption{Search trajectory of Gemini 3.6 Flash.}
\label{fig:trajectory-gemini}
\par\medskip
\centering
\includegraphics[width=0.96\textwidth,height=0.42\textheight,keepaspectratio]
{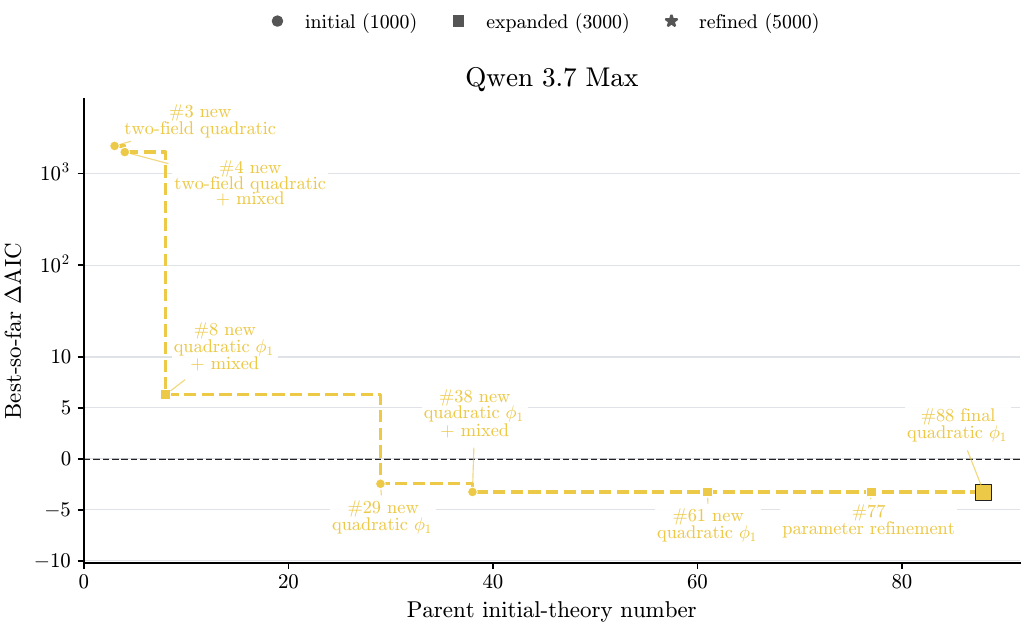}
\caption{Search trajectory of Qwen 3.7 Max.}
\label{fig:trajectory-qwen}
\end{figure*}

\begin{figure*}[p]
\centering
\includegraphics[width=0.96\textwidth,height=0.42\textheight,keepaspectratio]
{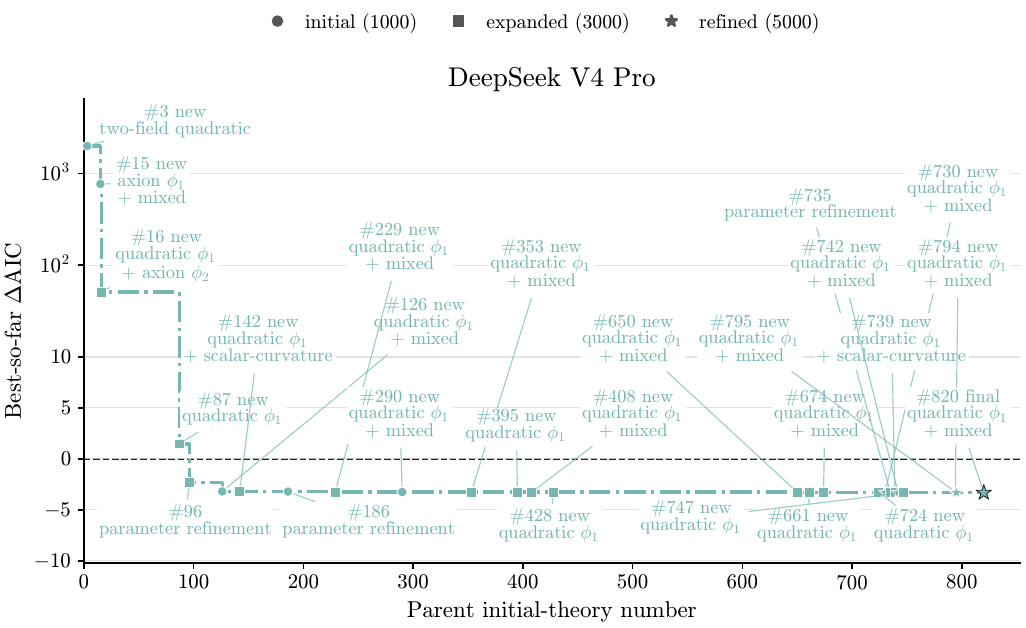}
\caption{Search trajectory of DeepSeek V4 Pro.}
\label{fig:trajectory-deepseek}
\par\medskip
\centering
\includegraphics[width=0.96\textwidth,height=0.42\textheight,keepaspectratio]
{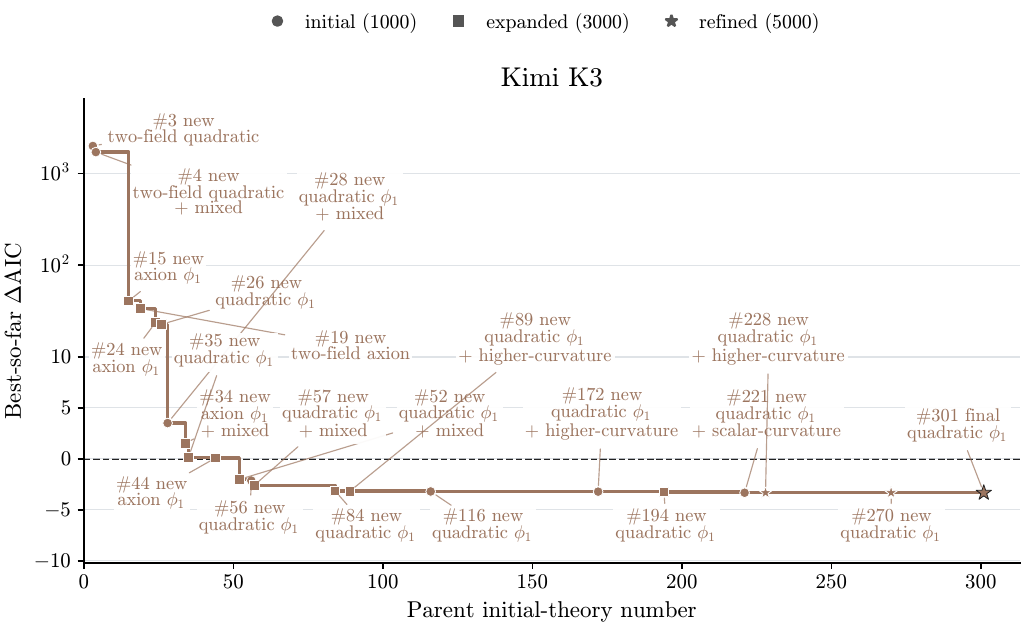}
\caption{Search trajectory of Kimi K3.}
\label{fig:trajectory-kimi}
\end{figure*}

\clearpage

\section{Grammar operator catalog}

\label{sec:Curvature invariants}

Tables~\ref{tab:operators-d2-d4}, \ref{tab:operators-d5}, and
\ref{tab:operators-d6} list the 141 candidate operators of mass
dimensions \(d=2,\ldots,6\) admitted by the truncated grammar,
including pure-curvature, scalar, and scalar--curvature terms. These
operators form the proposal dictionary rather than an independent
operator basis. Redundancies can affect how often equivalent theories
are proposed and hence the efficiency of the search. General
classifications of local curvature-polynomial bases are given in
Ref.~\cite{Fulling1992}.

\begin{table*}[t]
\caption{FLRW-distinct operator representatives of dimensions $d=2,3,4$.}
\label{tab:operators-d2-d4}
\centering
\footnotesize
\renewcommand{\arraystretch}{1.08}
\setlength{\tabcolsep}{4pt}
\begin{tabular*}{\textwidth}{@{\extracolsep{\fill}}llll@{}}
\hline\hline
ID & Operator & ID & Operator \\
\hline
$\mathcal{O}^{(2,\mathrm{S})}_{001}$ & $\phi_1^{2}$ &
$\mathcal{O}^{(4,\mathrm{S})}_{017}$ & $\nabla^{a}\phi_2\nabla_{a}\phi_2$ \\
$\mathcal{O}^{(2,\mathrm{S})}_{002}$ & $\phi_1\phi_2$ &
$\mathcal{O}^{(4,\mathrm{S})}_{018}$ & $\phi_1\Box\phi_1$ \\
$\mathcal{O}^{(2,\mathrm{S})}_{003}$ & $\phi_2^{2}$ &
$\mathcal{O}^{(4,\mathrm{S})}_{019}$ & $\phi_1\Box\phi_2$ \\
$\mathcal{O}^{(2,\mathrm{G})}_{004}$ & $R$ &
$\mathcal{O}^{(4,\mathrm{S})}_{020}$ & $\phi_1^{4}$ \\
$\mathcal{O}^{(3,\mathrm{S})}_{005}$ & $\Box\phi_1$ &
$\mathcal{O}^{(4,\mathrm{S})}_{021}$ & $\phi_1^{3}\phi_2$ \\
$\mathcal{O}^{(3,\mathrm{S})}_{006}$ & $\Box\phi_2$ &
$\mathcal{O}^{(4,\mathrm{S})}_{022}$ & $\phi_1^{2}\phi_2^{2}$ \\
$\mathcal{O}^{(3,\mathrm{S})}_{007}$ & $\phi_1^{3}$ &
$\mathcal{O}^{(4,\mathrm{NM})}_{023}$ & $\phi_1^{2}R$ \\
$\mathcal{O}^{(3,\mathrm{S})}_{008}$ & $\phi_1^{2}\phi_2$ &
$\mathcal{O}^{(4,\mathrm{S})}_{024}$ & $\phi_1\phi_2^{3}$ \\
$\mathcal{O}^{(3,\mathrm{S})}_{009}$ & $\phi_1\phi_2^{2}$ &
$\mathcal{O}^{(4,\mathrm{NM})}_{025}$ & $\phi_1\phi_2R$ \\
$\mathcal{O}^{(3,\mathrm{NM})}_{010}$ & $\phi_1R$ &
$\mathcal{O}^{(4,\mathrm{S})}_{026}$ & $\phi_2\Box\phi_2$ \\
$\mathcal{O}^{(3,\mathrm{S})}_{011}$ & $\phi_2^{3}$ &
$\mathcal{O}^{(4,\mathrm{S})}_{027}$ & $\phi_2^{4}$ \\
$\mathcal{O}^{(3,\mathrm{NM})}_{012}$ & $\phi_2R$ &
$\mathcal{O}^{(4,\mathrm{NM})}_{028}$ & $\phi_2^{2}R$ \\
$\mathcal{O}^{(4,\mathrm{S})}_{013}$ & $\Box\phi_1\,\phi_2$ &
$\mathcal{O}^{(4,\mathrm{G})}_{029}$ & $R^{2}$ \\
$\mathcal{O}^{(4,\mathrm{G})}_{014}$ & $\Box R$ &
$\mathcal{O}^{(4,\mathrm{G})}_{030}$ & $R^{ab}R_{ab}$ \\
$\mathcal{O}^{(4,\mathrm{S})}_{015}$ &
$\nabla^{a}\phi_1\nabla_{a}\phi_1$ &
$\mathcal{O}^{(4,\mathrm{G})}_{031}$ &
$R^{bcd}{}_{a}R^{a}{}_{bcd}$ \\
$\mathcal{O}^{(4,\mathrm{S})}_{016}$ &
$\nabla^{a}\phi_1\nabla_{a}\phi_2$ &
& \\
\hline\hline
\end{tabular*}
\end{table*}

\begin{table*}[t]
\caption{FLRW-distinct operator representatives of dimension $d=5$.}
\label{tab:operators-d5}
\centering
\footnotesize
\renewcommand{\arraystretch}{1.08}
\setlength{\tabcolsep}{4pt}
\begin{tabular*}{\textwidth}{@{\extracolsep{\fill}}llll@{}}
\hline\hline
ID & Operator & ID & Operator \\
\hline
$\mathcal{O}^{(5,\mathrm{S})}_{032}$ & $\Box\phi_1\,\phi_2^{2}$ &
$\mathcal{O}^{(5,\mathrm{S})}_{050}$ & $\phi_1^{3}\phi_2^{2}$ \\
$\mathcal{O}^{(5,\mathrm{NM})}_{033}$ & $\Box\phi_1R$ &
$\mathcal{O}^{(5,\mathrm{NM})}_{051}$ & $\phi_1^{3}R$ \\
$\mathcal{O}^{(5,\mathrm{NM})}_{034}$ &
$\nabla^{a}\nabla^{b}\phi_1R_{ab}$ &
$\mathcal{O}^{(5,\mathrm{S})}_{052}$ & $\phi_1^{2}\phi_2^{3}$ \\
$\mathcal{O}^{(5,\mathrm{NM})}_{035}$ & $\Box\phi_2R$ &
$\mathcal{O}^{(5,\mathrm{NM})}_{053}$ & $\phi_1^{2}\phi_2R$ \\
$\mathcal{O}^{(5,\mathrm{NM})}_{036}$ &
$\nabla^{a}\nabla^{b}\phi_2R_{ab}$ &
$\mathcal{O}^{(5,\mathrm{S})}_{054}$ & $\phi_1\phi_2\Box\phi_2$ \\
$\mathcal{O}^{(5,\mathrm{S})}_{037}$ &
$\nabla^{a}\phi_1\nabla_{a}\phi_1\,\phi_2$ &
$\mathcal{O}^{(5,\mathrm{S})}_{055}$ & $\phi_1\phi_2^{4}$ \\
$\mathcal{O}^{(5,\mathrm{NM})}_{038}$ &
$\nabla^{a}\phi_1\nabla_{a}R$ &
$\mathcal{O}^{(5,\mathrm{NM})}_{056}$ & $\phi_1\phi_2^{2}R$ \\
$\mathcal{O}^{(5,\mathrm{S})}_{039}$ &
$\nabla^{a}\phi_1\,\phi_2\nabla_{a}\phi_2$ &
$\mathcal{O}^{(5,\mathrm{NM})}_{057}$ & $\phi_1R^{2}$ \\
$\mathcal{O}^{(5,\mathrm{NM})}_{040}$ &
$\nabla^{a}\phi_2\nabla_{a}R$ &
$\mathcal{O}^{(5,\mathrm{NM})}_{058}$ & $\phi_1R^{ab}R_{ab}$ \\
$\mathcal{O}^{(5,\mathrm{S})}_{041}$ &
$\phi_1\Box\phi_1\,\phi_2$ &
$\mathcal{O}^{(5,\mathrm{NM})}_{059}$ &
$\phi_1R^{bcd}{}_{a}R^{a}{}_{bcd}$ \\
$\mathcal{O}^{(5,\mathrm{NM})}_{042}$ & $\phi_1\Box R$ &
$\mathcal{O}^{(5,\mathrm{NM})}_{060}$ & $\phi_2\Box R$ \\
$\mathcal{O}^{(5,\mathrm{S})}_{043}$ &
$\phi_1\nabla^{a}\phi_1\nabla_{a}\phi_1$ &
$\mathcal{O}^{(5,\mathrm{S})}_{061}$ &
$\phi_2\nabla^{a}\phi_2\nabla_{a}\phi_2$ \\
$\mathcal{O}^{(5,\mathrm{S})}_{044}$ &
$\phi_1\nabla^{a}\phi_1\nabla_{a}\phi_2$ &
$\mathcal{O}^{(5,\mathrm{S})}_{062}$ & $\phi_2^{2}\Box\phi_2$ \\
$\mathcal{O}^{(5,\mathrm{S})}_{045}$ &
$\phi_1\nabla^{a}\phi_2\nabla_{a}\phi_2$ &
$\mathcal{O}^{(5,\mathrm{S})}_{063}$ & $\phi_2^{5}$ \\
$\mathcal{O}^{(5,\mathrm{S})}_{046}$ & $\phi_1^{2}\Box\phi_1$ &
$\mathcal{O}^{(5,\mathrm{NM})}_{064}$ & $\phi_2^{3}R$ \\
$\mathcal{O}^{(5,\mathrm{S})}_{047}$ & $\phi_1^{2}\Box\phi_2$ &
$\mathcal{O}^{(5,\mathrm{NM})}_{065}$ & $\phi_2R^{2}$ \\
$\mathcal{O}^{(5,\mathrm{S})}_{048}$ & $\phi_1^{5}$ &
$\mathcal{O}^{(5,\mathrm{NM})}_{066}$ & $\phi_2R^{ab}R_{ab}$ \\
$\mathcal{O}^{(5,\mathrm{S})}_{049}$ & $\phi_1^{4}\phi_2$ &
$\mathcal{O}^{(5,\mathrm{NM})}_{067}$ &
$\phi_2R^{bcd}{}_{a}R^{a}{}_{bcd}$ \\
\hline\hline
\end{tabular*}
\end{table*}

\begin{table*}[t]
\caption{FLRW-distinct operator representatives of dimension $d=6$.}
\label{tab:operators-d6}
\centering
\footnotesize
\renewcommand{\arraystretch}{1.05}
\setlength{\tabcolsep}{3pt}
\begin{tabular*}{\textwidth}{@{\extracolsep{\fill}}llll@{}}
\hline\hline
ID & Operator & ID & Operator \\
\hline
$\mathcal{O}^{(6,\mathrm{S})}_{068}$ & $\Box\phi_1\,\phi_2^{3}$ &
$\mathcal{O}^{(6,\mathrm{NM})}_{105}$ & $\phi_1^{4}R$ \\
$\mathcal{O}^{(6,\mathrm{NM})}_{069}$ & $\Box\phi_1\,\phi_2R$ &
$\mathcal{O}^{(6,\mathrm{S})}_{106}$ & $\phi_1^{3}\phi_2^{3}$ \\
$\mathcal{O}^{(6,\mathrm{NM})}_{070}$ &
$\nabla^{a}\nabla^{b}\phi_1\,\phi_2R_{ab}$ &
$\mathcal{O}^{(6,\mathrm{NM})}_{107}$ & $\phi_1^{3}\phi_2R$ \\
$\mathcal{O}^{(6,\mathrm{G})}_{071}$ & $\Box R\,R^{c}{}_{c}$ &
$\mathcal{O}^{(6,\mathrm{S})}_{108}$ &
$\phi_1^{2}\phi_2\Box\phi_2$ \\
$\mathcal{O}^{(6,\mathrm{G})}_{072}$ &
$\nabla^{a}\nabla^{b}R\,R_{ab}$ &
$\mathcal{O}^{(6,\mathrm{S})}_{109}$ & $\phi_1^{2}\phi_2^{4}$ \\
$\mathcal{O}^{(6,\mathrm{S})}_{073}$ &
$\nabla^{a}\phi_1\nabla_{a}\phi_1\,\phi_2^{2}$ &
$\mathcal{O}^{(6,\mathrm{NM})}_{110}$ & $\phi_1^{2}\phi_2^{2}R$ \\
$\mathcal{O}^{(6,\mathrm{NM})}_{074}$ &
$\nabla^{a}\phi_1\nabla_{a}\phi_1R$ &
$\mathcal{O}^{(6,\mathrm{NM})}_{111}$ & $\phi_1^{2}R^{2}$ \\
$\mathcal{O}^{(6,\mathrm{NM})}_{075}$ &
$\nabla^{a}\phi_1\nabla^{b}\phi_1R_{ab}$ &
$\mathcal{O}^{(6,\mathrm{NM})}_{112}$ & $\phi_1^{2}R^{ab}R_{ab}$ \\
$\mathcal{O}^{(6,\mathrm{NM})}_{076}$ &
$\nabla^{a}\phi_1\nabla_{a}\phi_2R$ &
$\mathcal{O}^{(6,\mathrm{NM})}_{113}$ &
$\phi_1^{2}R^{bcd}{}_{a}R^{a}{}_{bcd}$ \\
$\mathcal{O}^{(6,\mathrm{NM})}_{077}$ &
$\nabla^{a}\phi_1\nabla^{b}\phi_2R_{ab}$ &
$\mathcal{O}^{(6,\mathrm{NM})}_{114}$ & $\phi_1\phi_2\Box R$ \\
$\mathcal{O}^{(6,\mathrm{NM})}_{078}$ &
$\nabla^{a}\phi_1\,\phi_2\nabla_{a}R$ &
$\mathcal{O}^{(6,\mathrm{S})}_{115}$ &
$\phi_1\phi_2\nabla^{a}\phi_2\nabla_{a}\phi_2$ \\
$\mathcal{O}^{(6,\mathrm{S})}_{079}$ &
$\nabla^{a}\phi_1\,\phi_2^{2}\nabla_{a}\phi_2$ &
$\mathcal{O}^{(6,\mathrm{S})}_{116}$ & $\phi_1\phi_2^{2}\Box\phi_2$ \\
$\mathcal{O}^{(6,\mathrm{NM})}_{080}$ &
$\nabla^{a}\phi_2\nabla_{a}\phi_2R$ &
$\mathcal{O}^{(6,\mathrm{S})}_{117}$ & $\phi_1\phi_2^{5}$ \\
$\mathcal{O}^{(6,\mathrm{NM})}_{081}$ &
$\nabla^{a}\phi_2\nabla^{b}\phi_2R_{ab}$ &
$\mathcal{O}^{(6,\mathrm{NM})}_{118}$ & $\phi_1\phi_2^{3}R$ \\
$\mathcal{O}^{(6,\mathrm{G})}_{082}$ & $\nabla^{a}R\nabla_{a}R$ &
$\mathcal{O}^{(6,\mathrm{NM})}_{119}$ & $\phi_1\phi_2R^{2}$ \\
$\mathcal{O}^{(6,\mathrm{G})}_{083}$ &
$\nabla^{c}R^{ab}\nabla_{c}R_{ab}$ &
$\mathcal{O}^{(6,\mathrm{NM})}_{120}$ & $\phi_1\phi_2R^{ab}R_{ab}$ \\
$\mathcal{O}^{(6,\mathrm{G})}_{084}$ &
$\nabla^{c}R^{ab}\nabla_{b}R_{ac}$ &
$\mathcal{O}^{(6,\mathrm{NM})}_{121}$ &
$\phi_1\phi_2R^{bcd}{}_{a}R^{a}{}_{bcd}$ \\
$\mathcal{O}^{(6,\mathrm{G})}_{085}$ &
$\nabla^{e}R^{bcd}{}_{a}\nabla_{e}R^{a}{}_{bcd}$ &
$\mathcal{O}^{(6,\mathrm{NM})}_{122}$ & $\phi_2\Box\phi_2R$ \\
$\mathcal{O}^{(6,\mathrm{S})}_{086}$ &
$\phi_1\Box\phi_1\,\phi_2^{2}$ &
$\mathcal{O}^{(6,\mathrm{NM})}_{123}$ &
$\phi_2\nabla^{a}\nabla^{b}\phi_2R_{ab}$ \\
$\mathcal{O}^{(6,\mathrm{NM})}_{087}$ & $\phi_1\Box\phi_1R$ &
$\mathcal{O}^{(6,\mathrm{NM})}_{124}$ &
$\phi_2\nabla^{a}\phi_2\nabla_{a}R$ \\
$\mathcal{O}^{(6,\mathrm{NM})}_{088}$ &
$\phi_1\nabla^{a}\nabla^{b}\phi_1R_{ab}$ &
$\mathcal{O}^{(6,\mathrm{NM})}_{125}$ & $\phi_2^{2}\Box R$ \\
$\mathcal{O}^{(6,\mathrm{NM})}_{089}$ & $\phi_1\Box\phi_2R$ &
$\mathcal{O}^{(6,\mathrm{S})}_{126}$ &
$\phi_2^{2}\nabla^{a}\phi_2\nabla_{a}\phi_2$ \\
$\mathcal{O}^{(6,\mathrm{NM})}_{090}$ &
$\phi_1\nabla^{a}\nabla^{b}\phi_2R_{ab}$ &
$\mathcal{O}^{(6,\mathrm{S})}_{127}$ & $\phi_2^{3}\Box\phi_2$ \\
$\mathcal{O}^{(6,\mathrm{S})}_{091}$ &
$\phi_1\nabla^{a}\phi_1\nabla_{a}\phi_1\,\phi_2$ &
$\mathcal{O}^{(6,\mathrm{S})}_{128}$ & $\phi_2^{6}$ \\
$\mathcal{O}^{(6,\mathrm{NM})}_{092}$ &
$\phi_1\nabla^{a}\phi_1\nabla_{a}R$ &
$\mathcal{O}^{(6,\mathrm{NM})}_{129}$ & $\phi_2^{4}R$ \\
$\mathcal{O}^{(6,\mathrm{S})}_{093}$ &
$\phi_1\nabla^{a}\phi_1\,\phi_2\nabla_{a}\phi_2$ &
$\mathcal{O}^{(6,\mathrm{NM})}_{130}$ & $\phi_2^{2}R^{2}$ \\
$\mathcal{O}^{(6,\mathrm{NM})}_{094}$ &
$\phi_1\nabla^{a}\phi_2\nabla_{a}R$ &
$\mathcal{O}^{(6,\mathrm{NM})}_{131}$ & $\phi_2^{2}R^{ab}R_{ab}$ \\
$\mathcal{O}^{(6,\mathrm{S})}_{095}$ &
$\phi_1^{2}\Box\phi_1\,\phi_2$ &
$\mathcal{O}^{(6,\mathrm{NM})}_{132}$ &
$\phi_2^{2}R^{bcd}{}_{a}R^{a}{}_{bcd}$ \\
$\mathcal{O}^{(6,\mathrm{NM})}_{096}$ & $\phi_1^{2}\Box R$ &
$\mathcal{O}^{(6,\mathrm{G})}_{133}$ & $R^{3}$ \\
$\mathcal{O}^{(6,\mathrm{S})}_{097}$ &
$\phi_1^{2}\nabla^{a}\phi_1\nabla_{a}\phi_1$ &
$\mathcal{O}^{(6,\mathrm{G})}_{134}$ & $R\,R^{ab}R_{ab}$ \\
$\mathcal{O}^{(6,\mathrm{S})}_{098}$ &
$\phi_1^{2}\nabla^{a}\phi_1\nabla_{a}\phi_2$ &
$\mathcal{O}^{(6,\mathrm{G})}_{135}$ &
$R\,R^{bcd}{}_{a}R^{a}{}_{bcd}$ \\
$\mathcal{O}^{(6,\mathrm{S})}_{099}$ &
$\phi_1^{2}\nabla^{a}\phi_2\nabla_{a}\phi_2$ &
$\mathcal{O}^{(6,\mathrm{G})}_{136}$ & $R^{ab}R^{d}{}_{a}R_{bd}$ \\
$\mathcal{O}^{(6,\mathrm{S})}_{100}$ & $\phi_1^{3}\Box\phi_1$ &
$\mathcal{O}^{(6,\mathrm{G})}_{137}$ &
$R^{b}{}_{a}R^{cd}R^{a}{}_{cbd}$ \\
$\mathcal{O}^{(6,\mathrm{S})}_{101}$ & $\phi_1^{3}\Box\phi_2$ &
$\mathcal{O}^{(6,\mathrm{G})}_{138}$ &
$R_{ab}R^{adef}R^{b}{}_{def}$ \\
$\mathcal{O}^{(6,\mathrm{S})}_{102}$ & $\phi_1^{6}$ &
$\mathcal{O}^{(6,\mathrm{G})}_{139}$ &
$R^{bd}{}_{ac}R^{agh}{}_{b}R^{c}{}_{dgh}$ \\
$\mathcal{O}^{(6,\mathrm{S})}_{103}$ & $\phi_1^{5}\phi_2$ &
$\mathcal{O}^{(6,\mathrm{G})}_{140}$ &
$R^{cd}{}_{ab}R^{afh}{}_{c}R^{b}{}_{fdh}$ \\
$\mathcal{O}^{(6,\mathrm{S})}_{104}$ & $\phi_1^{4}\phi_2^{2}$ &
$\mathcal{O}^{(6,\mathrm{G})}_{141}$ &
$R^{cd}{}_{ab}R^{afh}{}_{c}R^{b}{}_{hdf}$ \\
\hline\hline
\end{tabular*}
\end{table*}

\section{Scalaron extension and validation run}
\label{app:scalaron-extension}

In the previous pipeline, an algebraic nonlinear term such as $R^2$ was varied
directly with respect to the FLRW metric.  The resulting metric equation
contains derivatives above second order, so the derivative-order filter either
set its coefficient to zero or rejected the theory before it could reach
Builder and Solver.  This excluded the usual healthy scalaron representation
of $f(R)$ gravity~\cite{SotiriouFaraoni2010}.  We therefore added a scalaron stage before the order
filter.  For exact monomials in $R$, $\phi_1$, and $\phi_2$, it applies
\begin{equation}
\begin{split}
f(R,\phi_i)&\longrightarrow F(\chi,\phi_i)R-U(\chi,\phi_i),\\
F(\chi,\phi_i)&=\partial_\chi f(\chi,\phi_i),\\
U(\chi,\phi_i)&=\chi F(\chi,\phi_i)-f(\chi,\phi_i).
\end{split}
\label{eq:scalaron-legendre-transform}
\end{equation}
The auxiliary-field equation gives $\chi=R$ when $f_{RR}\ne0$.

\noindent\textit{Scope.}
The scalaron sector is restricted to the part of the dimension-six search
space of the form
\begin{equation}
f_{\rm alg}(R,\phi_1,\phi_2)
=\alpha R+
\sum_{\substack{n\ge2,\;p,q\ge0\\2n+p+q\le6}}
c_{npq}\Lambda^{2-(2n+p+q)}
R^n\phi_1^p\phi_2^q .
\label{eq:scalaron-scope}
\end{equation}
It therefore includes, for example, $R^2$, $R^3$, $\phi_iR^2$, and
$\phi_i\phi_jR^2$, but not higher-dimension powers outside the operator basis.
The search permits no matter scalar, either one of $\phi_1$ and $\phi_2$, or
both; active matter scalars have canonical kinetic terms and quadratic or
axion potentials.  Ricci- and Riemann-tensor contractions, $\Box R$, and
derivatives of curvature are not scalarized.  They may still coexist with
Eq.~\eqref{eq:scalaron-scope}, in which case they are preserved as residual
operators and tested by the original derivative-order filter.  Thus the
$f(R)$ rows below mean \emph{parent theories containing an eligible
$f(R,\phi_i)$ sector}; they are not restricted to pure $f(R)$ actions.

Scalarization is only the first step of the calculation.  The EOM stage varies
the transformed action with respect to the lapse, scale factor, matter
scalars, and $\chi$, and applies the derivative-order test to this coupled
system and to every residual operator.  Builder then introduces
\begin{equation}
\bar\chi=\frac{\chi}{H_0^2},
\qquad
q_\chi=\frac{d\bar\chi}{d\ln a},
\label{eq:scalaron-background-variables}
\end{equation}
and constructs the first-order background system for $H$, $\bar\chi$,
$q_\chi$, and any active matter scalars.  Solver samples $\bar\chi_0$, solves
the initial lapse constraint for $q_{\chi,0}$, integrates the background, and
monitors the constraint along the full trajectory.  It rejects a trajectory
if $F\le0$ or $f_{RR}\le0$ at any sampled redshift; for a pure $f(R)$ sector it
also requires positive scalaron mass squared~\cite{DeFeliceTsujikawa2010}.  Only surviving trajectories are
compared with DES-SN and DESI BAO data to obtain $\chi^2$ and AIC.  Thus the new
stage does not relax the original filter: it exposes the additional scalar
degree of freedom and then sends the equivalent theory through the complete
deterministic pipeline.  These remain background-level checks rather than a
full perturbative stability analysis.

We tested the extended pipeline in a cold-start run of 250 parent
theory proposals selected using GPT-5.6 Sol. The grammar-generated
candidates contain one to eight operators from the same dimension-six
grammar.  Each parent received 1000 initial parameter
samples, followed by the existing 3000- and 5000-sample stages when triggered.
The results below retain the best fit for each parent across these stages.

The mock diagnostic in Sec.~\ref{subsec:mock-lcdm250} and the
comparisons in Appendices~\ref{app:random-search}
and~\ref{app:memory-ablation} use the recorded
settings of this run or of the original benchmark. The implementations
used for the supplementary runs were archived separately, but no immutable
source snapshot is available for the historical LLM runs. Matching
therefore refers to the documented search settings, rather than a
verified identity of historical source versions.

\begin{table*}[t]
\centering
\caption{Physical-family summary of the 68 parent theories that achieved
$\chi^2_{\rm tot}<2000$.  The $f(R)$ rows use the scope in
Eq.~\eqref{eq:scalaron-scope}.  Standard quintessence denotes a reduced
single-field fit with no additional free correction coefficient.  Each
representative minimizes nominal AIC within its row; $Tn$ denotes
\texttt{task\_000$n$}.}
\label{tab:scalaron250-physical-summary}
\small
\setlength{\tabcolsep}{6pt}
\begin{ruledtabular}
\begin{tabular}{lrrcrrr}
Theory family
& $N_{\chi^2<2000}$
& Fraction
& Best parent
& $\chi^2_{\rm tot}$
& $k_{\rm nom}$
& ${\rm AIC}_{\rm nom}$ \\
\hline
$f(R)$, no matter scalar
& 27 & $39.7\%$ & T041 & 1649.0194 & 7 & 1663.0194 \\
$f(R)+\phi_1$
& 2 & $2.9\%$ & T190 & 1639.9789 & 10 & 1659.9789 \\
$f(R)+\phi_2$
& 7 & $10.3\%$ & T232 & 1660.1537 & 11 & 1682.1537 \\
$f(R)+\phi_1+\phi_2$
& 0 & $0.0\%$ & -- & -- & -- & -- \\
\textbf{All $f(R)$-containing theories}
& \textbf{36} & $\mathbf{52.9\%}$ & T190 & 1639.9789 & 10 & 1659.9789 \\
\hline
Quadratic quintessence
& 18 & $26.5\%$ & T248 & 1638.0349 & 5 & 1648.0349 \\
Axion quintessence
& 5 & $7.4\%$ & T048 & 1641.6697 & 6 & 1653.6697 \\
\textbf{All standard quintessence}
& \textbf{23} & $\mathbf{33.8\%}$ & T248 & 1638.0349 & 5 & 1648.0349 \\
\hline
Other scalar/curvature theories
& 9 & $13.2\%$ & T170 & 1638.0580 & 6 & 1650.0580 \\
\hline
\textbf{All selected parents}
& \textbf{68} & $\mathbf{100\%}$ & T248 & 1638.0349 & 5 & 1648.0349 \\
\end{tabular}
\end{ruledtabular}
\end{table*}

Table~\ref{tab:scalaron250-physical-summary} summarizes the 68 parents
($27.2\%$ of the 250 proposals) that reached $\chi^2_{\rm tot}<2000$.
Thirty-six contain a nonlinear $f(R)$ sector, including nine with a matter
scalar, while 23 reduce to standard single-field quintessence.  The remaining
nine retain additional scalar/curvature structure or more than one matter
scalar.  All 68 selected parents contain an Einstein--Hilbert anchor.  A further
28 parents have a finite fit with
$\chi^2_{\rm tot}\ge2000$, and 154 produced no finite sampled background.
Re-evaluation of every finite best-fit point reproduced the stored $\chi^2$ to
a maximum absolute difference of $1.7\times10^{-9}$.

Among the parents containing an eligible nonlinear $f(R,\phi_i)$
sector, T190 has the lowest nominal AIC. It illustrates how the scalaron
stage treats a theory with residual curvature operators. Before the
pre-filter, its gravity--scalar Lagrangian is
\begin{equation}
\begin{split}
\mathcal L_{\rm T190}=\Bigg\{M_p^2\Bigg[
&\alpha R+\frac{\beta}{\Lambda^2}R^2+\mathcal L_{\rm res}\Bigg]\\
&-\frac12\nabla_\mu\phi_1\nabla^\mu\phi_1
-\frac12m_1^2\phi_1^2\Bigg\},
\end{split}
\label{eq:t190-raw}
\end{equation}
where
\begin{equation}
\begin{split}
\mathcal L_{\rm res}={}&
\frac{c_3}{\Lambda^4}
R^{bd}{}_{ac}R^{agh}{}_{b}R^{c}{}_{dgh}
+\frac{c_4}{\Lambda^4}\nabla^\mu R\nabla_\mu R\\
&+\frac{c_5}{\Lambda^4}
\nabla^\mu\nabla^\nu R\,R_{\mu\nu}.
\end{split}
\label{eq:t190-residual}
\end{equation}
For $\beta\ne0$, the former metric-only order test did not pass the $R^2$
sector of this parent.  The new stage transforms only the first two curvature
terms,
\begin{equation}
\begin{split}
\mathcal L_{\rm T190}^{(\chi)}=\Bigg\{M_p^2\Bigg[
&\left(\alpha+\frac{2\beta\chi}{\Lambda^2}\right)R
-\frac{\beta\chi^2}{\Lambda^2}
+\mathcal L_{\rm res}\Bigg]\\
&-\frac12\nabla_\mu\phi_1\nabla^\mu\phi_1
-\frac12m_1^2\phi_1^2\Bigg\}.
\end{split}
\label{eq:t190-scalaron}
\end{equation}
Its auxiliary equation is
$2\beta(R-\chi)/\Lambda^2=0$, so the original action is recovered for
$\beta\ne0$.  The residual operators in Eq.~\eqref{eq:t190-residual} are not
scalarized and remain subject to the derivative-order relations.  The best
sample gives $\chi^2_{\rm tot}=1639.9789$ and, using the original nominal count
$k_{\rm nom}=10$, ${\rm AIC}_{\rm nom}=1659.9789$.

\subsection{Nominal and identifiable parameter counts}
\label{app:parameter-count-comparison}

The reported AIC values use the nominal count $k_{\rm nom}$ of free
parameters remaining after symbolic cancellations and Builder reduction.
This prescription is applied uniformly, but it may retain parameters
related by exact reparameterizations. We therefore examined how the
best-fit background responds to changes in these parameters.

\begin{samepage}
At each best-fit point, we estimated the background Jacobian
\[
J_{ai}=\frac{\partial\ln h(z_a)}{\partial\theta_i}
\]
by finite differences on the Solver grid, where $\theta_i$ denotes a
logarithmic or locally rescaled parameter coordinate.\par
\end{samepage}
\noindent
The initial step is $10^{-3}$ in these coordinates; the
routine uses one-sided differences or alternative step sizes when nearby
samples fail the Solver checks. Resolved columns with Euclidean norm
above $10^{-10}$ are normalized to unit norm, while smaller responses
are set to zero. Singular values of the resulting matrix $\widetilde J$
above $10^{-4}$ times its largest singular value contribute to the
numerical rank. The diagnostic count is
\begin{equation}
k_{\rm id}={\rm rank}(\widetilde J)
           +N_{\rm unresolved}+N_{\rm profiled}.
\end{equation}
Here $N_{\rm unresolved}$ counts parameter directions for which the
finite-difference calculation cannot obtain a valid background response;
each is retained as one fit direction. The profiled SN offset and BAO
scale do not enter $h(z)$ and are counted separately, giving
$N_{\rm profiled}=2$.

We interpreted the numerically null directions by inspecting the
parameters involved. Directions involving only the absolute scale or
the cutoff and curvature coefficients were flagged as possible
reparameterization redundancies. Directions involving scalar masses,
field values, matter density, or corrections inactive at the best-fit
point were treated as local weak identification; models with both
types were classified as mixed. These classifications guide further
symbolic checks and do not establish exact equivalences by themselves.

Of the 96 parents with finite fits, 14 retained full nominal rank,
35 were flagged as candidates for structural redundancy, 20 showed
only local weak identification, 25 were mixed, and two calculations
were incomplete. We retain $k_{\rm nom}$ for AIC comparisons until an
explicit symbolic reduction establishes that a parameter is redundant.
The best result in Table~\ref{tab:scalaron250-physical-summary} remains
the quadratic-quintessence model T248, with $k_{\rm nom}=k_{\rm id}=5$.

\section{Acceptance and rejection diagnostics}
\label{app:acceptance-diagnostics}

We examined the archived initial-evaluation records for the eight
observed-data benchmark runs in Table~\ref{tab:agent-benchmark}.
Each run contributes 1000 completed parent proposals. Expanded and
refined evaluations are excluded, and repeated theory structures remain
separate proposals, as in the benchmark acceptance denominator.
A proposal is accepted when its initial evaluation returns at least
one parameter sample that passes the implemented background and
numerical checks and yields a finite likelihood. This criterion imposes
no relative AIC requirement or goodness-of-fit threshold.

Of the 8000 initial evaluations, 2570 are accepted and 5430 are
unsuccessful. The recorded terminal stages for the latter are Builder
in 5 cases and Solver in 5425 cases; none terminates at EOM within
this set of completed evaluations. Of the Solver failures, 5415 report
no accepted ranking row and 10 report a Solver-process failure.
The terminal stage identifies where the failure is reported, not
necessarily its cause: an unusable symbolic model can also lead to
an empty Solver ranking.

Table~\ref{tab:rejection-diagnostics} groups the explicit rejection
messages retained in the initial-evaluation records. Constraint drift
and failure to find acceptable real roots for the initial scalar
velocities are the most frequently recorded diagnostics. Other
evaluations fail because the sampled no-scalar configuration does not
satisfy the initial constraint, the reduced equations do not determine
the scalar accelerations, symbolic construction exceeds its time limit,
or the numerical trajectory fails an integration or admissibility check.

\begin{table}[ht]
\centering
\caption{Diagnostics in the retained rejection summaries of the
observed-data benchmark. Each initial proposal is counted once per
listed category. Percentages use all 5430 unsuccessful initial
evaluations. Categories overlap because different parameter samples
from one proposal can fail for different reasons.}
\label{tab:rejection-diagnostics}
\small
\setlength{\tabcolsep}{3pt}
\renewcommand{\arraystretch}{1.2}
\begin{tabular}{lrr}
\hline
Recorded diagnostic & Proposals & (\%) \\
\hline
\shortstack[l]{Constraint drift exceeds\\tolerance}
  & 1859 & 34.24 \\
\shortstack[l]{No acceptable initial\\scalar-velocity root}
  & 1824 & 33.59 \\
\shortstack[l]{Initial constraint fails\\without scalars}
  & 815 & 15.01 \\
\shortstack[l]{Scalar acceleration\\not determined}
  & 516 & 9.50 \\
Symbolic-build timeout & 395 & 7.27 \\
\shortstack[l]{Singular or excessive\\initial derivatives}
  & 274 & 5.05 \\
Numerical integration fails & 250 & 4.60 \\
\shortstack[l]{Expansion rate exceeds\\its bound}
  & 217 & 4.00 \\
\shortstack[l]{Scalar amplitude or velocity\\exceeds its bound}
  & 112 & 2.06 \\
EFT cutoff violated & 16 & 0.29 \\
\hline
\end{tabular}
\end{table}

These counts use the explicit rejection messages rather than the
coarser stored failure tags. Rejection summaries are present for 5275
of the unsuccessful evaluations; the remaining 155 have no retained
reason list and are included only in the stage counts above. Each
summary retains at most eight reason entries, so the table describes
documented occurrences and is not an exhaustive breakdown of all
rejected parameter samples. In particular, a numerical failure or
timeout does not establish a physical inconsistency of the full theory.
The acceptance rate measures the yield of the implemented background
search with its stated priors, parameter budget, and tolerances.

\section{Random-search baseline}
\label{app:random-search}

We performed a separate search of 1000 parent theory proposals on the
observed DES-SN and DESI-DR2 BAO data to assess the performance of fixed
random proposals within the prescribed grammar. The four scalar
configurations were sampled with equal probability, followed by a
uniform choice of one to eight generated operators, a uniform choice
among the allowed potential forms, and a uniformly sampled subset of
operators compatible with the selected scalar configuration. This
defines a distribution over grammar representations; equivalent
Lagrangians need not have equal sampling probability. The usual
duplicate limits were retained, and no preference for an
Einstein--Hilbert term was imposed at proposal time.

Continuous parameters were drawn using the fixed default ranges in
Table~\ref{tab:parameter-priors}. The prescribed mixture of direct
coefficient draws and draws based on the relative size of curvature
corrections was also retained, with fixed weights of 0.10, 0.30, 0.30,
and 0.30 for the baseline, weak, moderate, and active-probe components.
Neither theory selection nor parameter sampling used previous fit
outcomes, learned priors, or stored successful parameter vectors.
Constraint-solved quantities and the profiled SN and BAO nuisance
parameters were treated as in the LLM-guided runs. Every proposal
underwent the grammar, scalaron, background derivative-order, Builder,
and Solver checks with the prescribed physical and numerical cuts.

Each parent received an initial allocation of 1000 parameter samples.
An initial accepted fit with $\chiT<2000$ triggered a further 3000-sample
scan of that structure, and an expanded fit with $\chiT<1638$ would
trigger a further 5000-sample scan. These within-theory gates did not
alter the priors for subsequent proposals. Eleven expanded scans and
no refined scans were triggered, giving 1011 validation blocks for
1000 parents. Acceptance is measured on the initial evaluations,
whereas the reported minimum AIC includes all triggered scans.
Sample allocations do not imply that every vector reaches likelihood
evaluation, since earlier validation stages can reject a proposal.

\begin{table}[ht]
\centering
\caption{Fixed-prior random search on the observed data. Differences
use the reported flat-$\Lambda$CDM reference of
Sec.~\ref{subsec:aic-objective}, with $k_{\Lambda{\rm CDM}}=3$ and $N=1833$.}
\label{tab:random-search}
\small
\begin{tabular}{lr}
\hline
Quantity & Result \\
\hline
Parent proposals & 1000 \\
Accepted initial proposals & 103 (10.3\%) \\
Expanded / refined scans & 11 / 0 \\
Best $\chiT$ & 1645.214 \\
$k$ of the minimum-AIC fit & 5 \\
Minimum AIC & 1655.214 \\
$\Delta\mathrm{AIC}$ & $+3.934$ \\
$\Delta\mathrm{BIC}$ of the same fit & $+14.961$ \\
Parents with $\Delta\mathrm{AIC}<0$ & 0 \\
\hline
\end{tabular}
\end{table}

The results are summarized in Table~\ref{tab:random-search}. The
minimum-AIC candidate has $\chi^2=1645.214$, close to the reported
$\Lambda$CDM reference value of 1645.28. Its two additional nominal fit parameters therefore
outweigh this small improvement under both AIC and BIC. The random
search can reach a competitive background fit, although its best AIC
and initial acceptance fraction are less favorable than those of the
recorded LLM-guided searches.

The random seed was 20260607, and the comparison is subject to the
source-version limitation described in Appendix~\ref{app:scalaron-extension}.
Equal parent budgets need not give equal numbers of numerical evaluations,
since additional scans depend on each proposal's fit. This single run
compares fixed random proposals with the combined LLM-guided and adaptive
search procedure. Appendix~\ref{app:memory-ablation} separately tests
current-run memory with 250 parents per setting. Repeated runs and
further component ablations would be needed to determine how
consistently each part of the procedure improves the search.

\section{Current-run memory ablation}
\label{app:memory-ablation}

We compared two GPT-5.6 Sol searches on the observed DES-SN and
DESI-DR2 BAO data, each with 250 parent proposals. The memory-enabled
reference is the scalaron-extension run described in
Appendix~\ref{app:scalaron-extension}. The second run retained LLM
candidate selection but disabled the use of earlier validation results
in both theory proposals and parameter sampling. In
particular, neither the learned structure prior, the marginal
parameter prior, nor the structure-conditioned memory of successful
parameter vectors was available. The fixed default priors, candidate
construction procedure, and mechanical duplicate limit were retained.
Results were still logged, but were not used to guide later proposals.
The LLM weights were unchanged in both runs.

The comparison used the same 141-operator grammar and observed-data
likelihood, with the recorded parameter ranges, validation cuts, and
1000/3000/5000 sample-allocation rules. Both searches used eight
asynchronous slots and numerical seed 20260607. An initial accepted
fit with $\chiT<2000$ triggered an additional 3000-sample scan, and
an expanded fit with $\chiT<1638$ triggered a further 5000-sample
scan. All parameter draws in the memory-disabled run used fixed
priors, including those in the additional scans. Every candidate
underwent the full symbolic and numerical validation pipeline.

\begin{table}[ht]
\centering
\caption{GPT-5.6 Sol searches with and without current-run memory
on the observed data. Each column represents one search.
Information criteria use the nominal parameter count and
$N=1833$; differences are relative to the same flat-$\Lambda$CDM
reference. The best-fit rows refer to the minimum-AIC candidate
across all initial and triggered additional scans.}
\label{tab:memory-ablation}
\small
\setlength{\tabcolsep}{4pt}
\begin{tabular}{lrr}
\hline
Quantity & Memory on & Memory off \\
\hline
Parent proposals & 250 & 250 \\
Accepted initial proposals & 96 & 99 \\
Initial acceptance (\%) & 38.4 & 39.6 \\
Validation blocks & 318 & 290 \\
Best-fit $\chiT$ & 1638.035 & 1642.025 \\
$k$ of the best fit & 5 & 5 \\
Minimum AIC & 1648.035 & 1652.025 \\
$\Delta\mathrm{AIC}$ & $-3.245$ & $+0.745$ \\
$\Delta\mathrm{BIC}$ & $+7.782$ & $+11.772$ \\
\hline
\end{tabular}
\end{table}

Table~\ref{tab:memory-ablation} shows that the two runs have similar
initial acceptance fractions, but the memory-enabled run reaches a
minimum AIC lower by 3.990. Both best fits have $k=5$, so the difference
comes entirely from $\chiT$. The memory-enabled candidate improves on
the reference AIC, whereas the memory-disabled candidate does not;
BIC favors the $\Lambda$CDM reference in both cases. These comparisons
describe the search outcomes and do not establish a departure from
$\Lambda$CDM.

The 290 validation records from the memory-disabled run confirm that
LLM selection was retained and that neither validation feedback nor
learned parameter vectors entered later proposals. The memory-enabled
run contains 318 validation blocks. The parent budgets and rules for
additional scans therefore match, but the realized numerical work
differs. Together with the source-version limitation described in
Appendix~\ref{app:scalaron-extension}, this restricts the comparison to
outcomes under the recorded search protocol, rather than efficiency
at fixed evaluation count or wall time.

The experiment tests feedback in theory selection and parameter
sampling together. With one run per setting, it does not separate
their contributions or measure the variability of the 3.990 AIC
difference; both questions require further runs.

\linespread{0.98}\selectfont
\bibliographystyle{apsrev4-2}
\bibliography{main}

\end{document}